\documentclass[11pt,preprint]{aastex}

\usepackage{apjfonts}
\usepackage{epsfig}
\usepackage{graphicx}
\usepackage{url}
\usepackage{natbib}
\usepackage{float}

\shorttitle{The Orbit of the Cetus Polar Stream}
\shortauthors{Yam et al.}

\begin{document}

\title{Update on the Cetus Polar Stream and its Progenitor}

\author{
William Yam\altaffilmark{1},
Jeffrey L. Carlin\altaffilmark{1},
Heidi Jo Newberg\altaffilmark{1},
Julie Dumas\altaffilmark{1},
Erin O'Malley\altaffilmark{1,2},
Matthew Newby\altaffilmark{1}
Charles Martin\altaffilmark{1}
}

\altaffiltext{1}{Department of Physics, Applied Physics and Astronomy, Rensselaer Polytechnic Institute, Troy, NY 12180, USA; carlij@rpi.edu}
\altaffiltext{2}{Department of Physics, Siena College}

\begin{abstract}

We trace the Cetus Polar Stream (CPS) with blue horizontal branch (BHB) and red giant stars (RGBs) from Data Release 8 of the Sloan Digital Sky 
Survey (SDSS DR8). Using a larger dataset than was available previously, we are able to refine the measured distance and velocity to this tidal debris 
star stream in the south Galactic cap. Assuming the tidal debris traces the progenitor's orbit, we fit an orbit to the CPS and find that the stream is 
confined between $\sim24-36$~kpc on a rather polar orbit inclined $87\arcdeg$ to the Galactic plane. The eccentricity of the orbit is 0.20, and the 
period $\sim700$~Myr. If we instead matched $N$-body simulations to the observed tidal debris, these orbital parameters would change by 10\% or 
less. The CPS stars travel in the opposite direction to those from the Sagittarius tidal stream in the same region of the sky. Through $N$-body models of 
satellites on the best-fitting orbit, and assuming that mass follows light, we show that the stream width, line-of-sight depth, and velocity dispersion imply 
a progenitor of $\gtrsim 10^8 M_{\Sun}$.  However, the density of stars along the stream requires either a disruption time on the order of one orbit, or a 
stellar population that is more centrally concentrated than the dark matter.  We suggest that an ultra-faint dwarf galaxy progenitor could reproduce a 
large stream width and velocity dispersion without requiring a very recent deflection of the progenitor into its current orbit.  We find that most Cetus stars 
have metallicities of $-2.5 <$ [Fe/H] $< -2.0$, similar to the observed metallicities of the ultra-faint dwarfs.  Our simulations suggest that the parameters 
of the dwarf galaxy progenitors, including their dark matter content, could be constrained by observations of their tidal tails through comparison of the 
debris with $N$-body simulations.

\end{abstract}


\section{Introduction}

Our current view of galaxy formation on Milky Way (MW) scales,
initially put forward by \citet{sz78}, is based on the idea of
hierarchical merging, the gradual agglomeration of tidally disrupted
dwarf galaxies and globular clusters onto larger host galaxies. The
prevailing $\Lambda$-cold dark matter model of structure formation
predicts frequent satellite disruption events continuing even to late
times (e.g., \citealt{j98a, mgg+99, ans+03a, bj05}). With the advent
of large-scale photometric (and, to a lesser extent, spectroscopic)
surveys covering large volumes of the Milky Way, the Galactic halo has
been revealed to be coursed with remnant streams from tidally
disrupted late-infalling satellites (e.g., Sagittarius:
\citealt{iil+01a, msw+03}; the Monoceros/Anticenter Stream complex:
\citealt{nyr+02, iil+03, yng+03, g06a, lnc+12}; Cetus Polar Stream:
\citealt{nyw09}; Virgo substructure: \citealt{vza+01, nyr+02, dzv+06,
cyc+12}; and various other SDSS streams: \citealt{bze+06, g06, gd06,
gd06a, gj06, bei+07, g09}; for a summary of known Milky Way stellar
streams, see \citealt{g10}).  The stars in tidal streams are extremely
sensitive probes of the underlying Galactic gravitational potential in
which they orbit; thus the measurement of kinematics in a number of
tidal streams traversing different regions of the Galaxy is an
essential tool to be used in mapping the Galactic (dark matter)
halo. This has been done for a handful of streams, the most prominent
of which are the extensively-studied tidal streams emanating from the
Sagittarius (Sgr) dwarf spheroidal galaxy.  The Sgr streams have been
used to argue for a MW dark matter halo that is nearly-spherical
(e.g., \citealt{ili+01, fbe+06}), oblate (e.g., \citealt{jlm05,
mpj+07}), prolate (e.g., \citealt{h04}), and finally triaxial
\citep{lmj09, lm10a}. \citet{krh10} used orbits fit to kinematical
data of stars in the "GD-1" stream (originally discovered by
\citealt{gd06}) to show that the Galactic potential is slightly oblate
within the narrow range of Galactic radii probed by the GD-1 stream.

In this work, we focus on the Cetus Polar Stream (CPS). This substructure was originally noted by \citet{ynj+09} as a low-metallicity group of blue
horizontal branch (BHB) stars near the Sgr trailing tidal tail among
data from the Sloan Extension for Galactic Understanding and
Exploration (SEGUE). The stream was found to be spatially coincident
with and at a similar distance to the trailing tidal tail of the
Sagittarius dwarf spheroidal near the south Galactic cap. The authors
noted that the stream is more metal poor than Sgr, with [Fe/H] $\sim$
-2.0, and has a different Galactocentric radial velocity trend than
Sgr stars in the same region of sky.

\citet{nyw09} used data from the Sloan Digital Sky Survey (SDSS) Data Release 7 (DR7) to confirm 
this discovery of a new stream in the south Galactic cap. The authors
showed that this new stream of BHB stars crosses the Sagittarius
trailing tidal tail at $(l,b) \sim (140\arcdeg, -70\arcdeg)$, but is
separated from Sgr by about $30\arcdeg$ in Galactic longitude at $b \sim
-30\arcdeg$. Because this newfound stream is located mostly in the
constellation Cetus and is roughly distributed along constant Galactic
longitude (i.e., the orbit is nearly polar), \citet{nyw09} dubbed it
the Cetus Polar Stream. A slight gradient in the distance to the
stream was detected, from $\sim 36$ kpc at $b \sim -71\arcdeg$ to $
\sim 30$ kpc at $b \sim -46\arcdeg$, where the distance nearer the South Galactic Pole places the 
CPS at approximately the same distance as the Sgr stream at that
position. The authors examined BHB, red giant branch (RGB), and lower
RGB (LRGB) stars identified by stellar parameters in the SEGUE
spectroscopic database, and found a mean metallicity of [Fe/H] = -2.1
for the CPS. The ratio of blue straggler (BS) to BHB stars was shown
to be much higher in Sgr than in the CPS, suggesting that most of the
BHB stars in this region of the sky may be associated with Cetus
rather than the Sgr stream, as previous studies had assumed. From the
limited SDSS data available in this region of the sky, this work
concluded that the spatial distribution of the CPS is noticeably
different than Sgr in the Southern Cap. Indeed, it was pointed out
that this solved a mystery seen by
\citet{ynk+00}, where the BHB stars in that earlier work did not seem to spatially coincide with 
the Sgr blue stragglers at $\sim2$ magnitudes fainter in the same SDSS
stripe; the majority of these BHB stars must not be Sgr members, but
CPS debris instead. The velocity signature of low- metallicity stars
clearly differs from the Sgr velocity trend in this region of the sky,
as was seen in the
\citet{ynj+09} study, further differentiating the CPS from Sgr. The velocity trend, positions, and 
distance estimates were used by \citet{nyw09} to fit an orbit to the
CPS. However, because SDSS data available at the time only covered
narrow stripes in the south Galactic cap, the orbit derived by Newberg
et al. in this study was rather uncertain.

In a study of the portion of the Sagittarius trailing stream in the
Southern Galactic hemisphere using SDSS Data Release 8 (DR8) data,
\citet{kbe+12} noted that the BS and BHB stars in the region near the
Sgr stream are not at a constant magnitude offset from each other as a
function of position, as would be expected if they were part of the
same population. The magnitude difference between color-selected blue
stragglers and BHBs is $\sim2$ magnitudes at $\Lambda \sim 130\arcdeg$
(where $\Lambda$ is longitude in a coordinate system rotated into the
Sagittarius orbital plane, such that larger $\Lambda$ is increasingly
further from the Sgr core along the trailing tail; \citealt{msw+03}),
and decreases at lower Sgr longitudes. Thus, the BHBs at $\Lambda $
well away from $130\arcdeg$ must not be associated with
Sagittarius. The distance modulus of these Cetus BHBs shows a clear
trend over more than $40\arcdeg$ along the stream. The authors show
that BS and BHB stars with SDSS DR8 spectra enable kinematical
separation of Sgr and the CPS, as had already been seen in the
\citet{nyw09} data from DR7. The kinematics of Cetus are suggested by
\citet{kbe+12} to imply a counter-rotating orbit with respect to Sgr.

In this paper we follow up on the work by \citet{nyw09} using the
extensive new imaging data made available in SDSS DR8, which now
provides complete coverage of much of the south Galactic cap
region. Additional spectra are available in DR8 as well, though they
are limited to the DR7 footprint consisting of a few SDSS stripes
only. The goal of this work is to measure the distances, positions,
and velocities of CPS stars along the stream to sufficient accuracies
that we can derive a reliable orbit for the Cetus Polar Stream. This
will enable new constraints on the shape and strength of the Milky
Way gravitational potential over the regions probed by Cetus, as well
as illuminating another of the merger events in the hierarchical
merging history of the Galactic halo. 
We then use this orbit to generate $N$-body models of the stream, 
and show that a $\sim10^8 M_{\Sun}$ satellite is required to reproduce the kinematics of the stream, but the distribution of stars along the stream requires either a short disruption time or a dwarf galaxy in which mass does not follow light. We argue that the CPS progenitor was likely a dark matter-dominated dwarf galaxy similar to the ultra-faint dwarfs in the Milky Way.

\section{Finding CPS from Photometrically Selected BHB stars}

Because blue horizontal branch (BHB) stars are prevalent in the CPS
\citep{nyw09,kbe+12}, we first select this population of stars in the
larger SDSS DR8 dataset. 
BHBs are
color-selected with $-0.25 < (g-r)_0 < -0.05$ and $0.8 < (u-g)_0 <
1.5$, where the latter cut removes most of the QSOs from the sample.
Here and throughout this paper, we use the subscript ``$_0$'' to
indicate that the magnitudes have been corrected by the \citet{sfd98}
extinction maps, as implemented in SDSS DR8.
BHB stars were selected in the Galactic
longitude range $120\arcdeg < l < 165\arcdeg$, bracketing the nearly
constant Galactic longitude of $l \sim 143\arcdeg$ found by
\citet{nyw09} for the CPS in the south Galactic cap.    We further separate
stars within the initial color selections that are more likely to be
higher surface gravity blue straggler (BS) stars from those more
likely to be lower surface gravity BHBs using the $ugr$ color-color
cuts outlined in \citet{ynk+00}.

\begin{figure}[!t]
\plotone{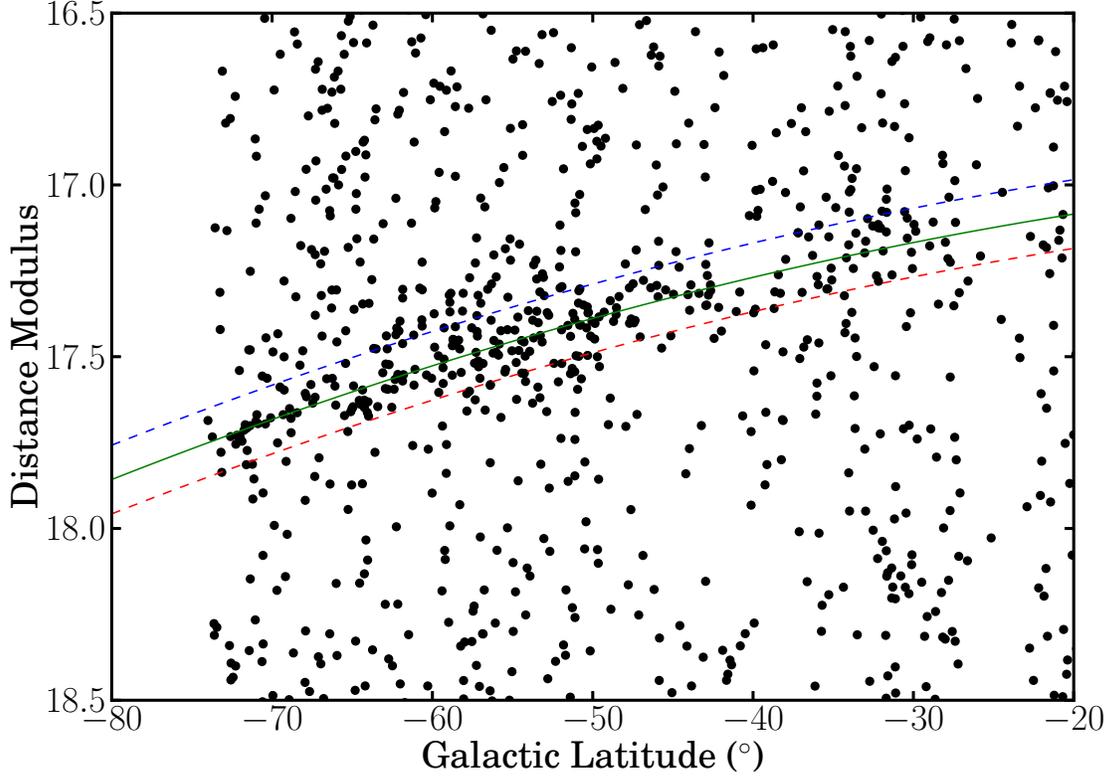}
\caption{Distance modulus to BHB stars, calculated by applying Equation~7 of \citet{dbe11} to find $M_{g, BHB}$ as a function of $(g-r)_0$ color, as a function of Galactic latitude. Because the stream is at roughly constant longitude, latitude corresponds approximately to angle along the
stream.  The BHBs were selected with $120\arcdeg < l < 165\arcdeg$,
$-0.25<(g-r)_0<-0.05$, $0.8 < (u-g)_0 < 1.5$, and using the $ugr$
color-color selection found by \citet{ynk+00} to favor low surface
gravity BHB stars while reducing the population of higher surface
gravity BS stars. We fit a quadratic function to the clear overdensity
of BHB stars and found $(g_0-M_{g_0}) = 16.974 - (0.003710 \times b) +
(0.000091686 \times b^2)$.  The solid green line shows this
trendline. Upper and lower bounds for selecting CPS BHB candidates are
shown as blue and red dashed lines 0.1 magnitudes above and below this
trendline.  The gap around a latitude of $b \sim -40\arcdeg$ is a
result of the fact that the SDSS footprint does not cover the full
Galactic longitude at this latitude (see Figure~\ref{lbppb}).  The
data were cut off at an upper bound of $b=-20\arcdeg$ because the data
at regions nearer the Galactic plane include primarily disk stars. The ``blob'' at $b\sim-30\arcdeg$, $D_{\rm mod} \sim 18.2$ is actually made up of extremely luminous blue supergiants in the galaxy M33, and not BHB stars as assumed when deriving the distance moduli.}
\label{gbpb}
\end{figure}

The extinction-corrected apparent magnitudes of the BHB stars are
converted to absolute magnitude, $M_{g_0}$, using the relation between
absolute $g-$band magnitude and $(g-r)_0$ color for BHB stars given by
Equation~7 of \citet{dbe11}. The distance modulus for each BHB star
then constitutes the difference between the apparent and absolute
magnitudes.  The distance moduli of the color-color selected BHB stars
are shown as a function of Galactic latitude in Figure~\ref{gbpb}.
Because the stream is extended along nearly constant Galactic
longitude, Galactic latitude is nearly the same as angular distance
along the stream.  In this graph there is a clear concentration of BHB
stars with an approximately linear relationship between distance
modulus (which is between $0.45 \lesssim M_{g_0} \lesssim 0.7$
magnitudes offset from the measured apparent magnitudes of these
stars, depending on color) and Galactic latitude, as was found in
\citet{nyw09} and \citet{kbe+12}. However, there is a hint that the
relationship is not quite linear.  Therefore, we chose to fit a
parabola to the BHB data between $17.0 < g_0-M_{g_0} < 18.0$ and $b <
-30\arcdeg$, iteratively rejecting outliers (beginning with $3\sigma$
rejection, then reducing this to $2\sigma$, and finally $1.5\sigma$)
until the fit converged to the one overlaid in Figure~\ref{gbpb} as a
solid line. This fit is given by

\begin{equation}
D_{mod} (b) \equiv (g_0-M_{g_0}) = 16.974 - (3.710\times10^{-3} \times b) + (9.1686\times10^{-5} \times b^2).
\label{dm_eqn}
\end{equation}

\noindent The dashed lines in the figure show $\pm0.1$-magnitude ranges about the fit for distance modulus as a function of latitude. 

\begin{figure}[!t]
\plotone{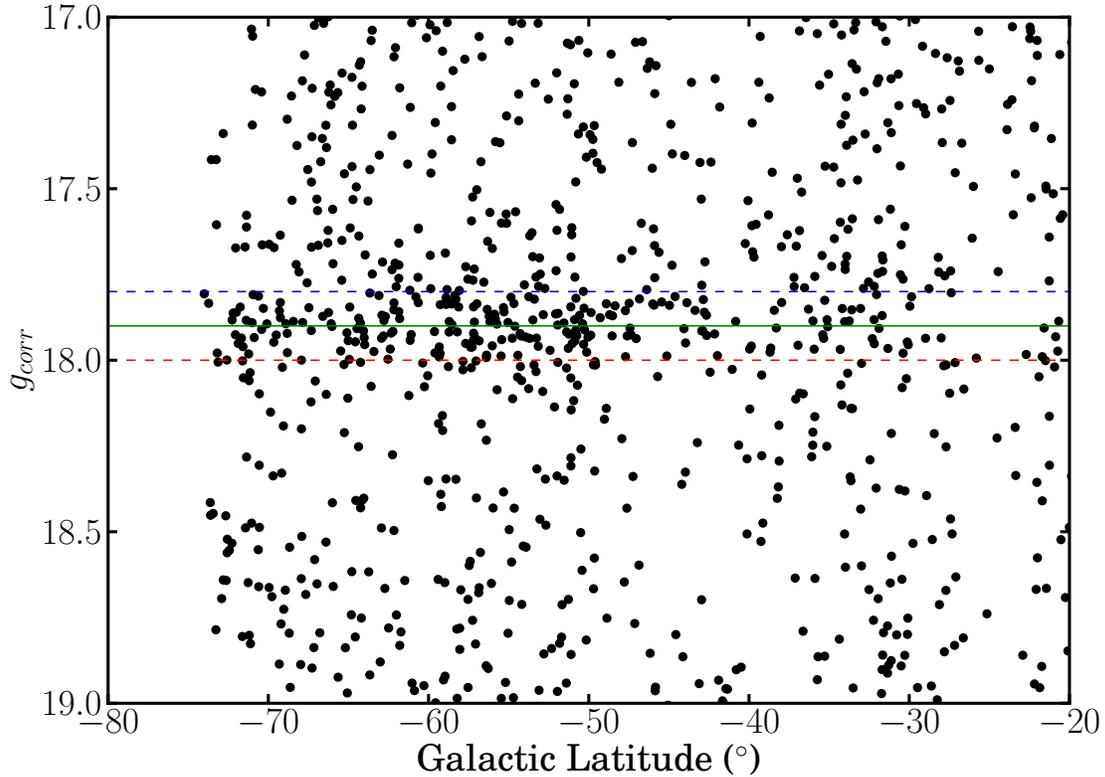}
\caption{Corrected magnitudes for BHB stars from Figure~\ref{gbpb} as a function of Galactic latitude. These are defined as $g_{corr} (b)\equiv g_0 - D_{mod} (b) + 17.389$, using the relationships given in
Equations~\ref{dm_eqn}~and~\ref{gcorr_eqn}. This corrects all of the
BHB star magnitudes for the distance modulus trend of CPS stars, placing them at
the CPS distance as measured at $b = -50\arcdeg$. Note that the
density of CPS BHB stars seems to drop off at latitudes above
$b>-40\arcdeg$ (i.e., closer to the disk).}
\label{gmodbpb}
\end{figure}

Figure~\ref{gmodbpb} shows the same data as the previous figure, but
with apparent magnitude corrected as a function of Galactic latitude
for the trend fit in Figure~\ref{gbpb} to place all stars at the BHB
distance corresponding to CPS stars at $b = -50\arcdeg$. These new
``corrected'' magnitudes, $g_{corr}$, are defined as:

\begin{equation}
g_{corr} (b) = g_0 - D_{mod} (b) + 17.389, 
\label{gcorr_eqn}
\end{equation}

\noindent where $D_{mod} (b)$ is the distance modulus fit as a function of latitude given by Equation~\ref{dm_eqn}, and 17.389 is the distance modulus of a star on the polynomial fit at $b = 
-50\arcdeg$.  The CPS stars cluster tightly about the fit in
Figure~\ref{gmodbpb} at high latitudes ($b < -40\arcdeg$), then drop
off near the Galactic plane (i.e., at $b > -40\arcdeg$). This may be a
real effect of the CPS BHB star density dropping off as one moves away
from the Galactic pole along the stream (though also partly due to the non-uniform coverage of SDSS; see Figure~\ref{lbppb} and further discussion in Sections~\ref{sec:position} and \ref{sec:nbody}).

\begin{figure}[!t]
\includegraphics[width=0.75\textwidth]{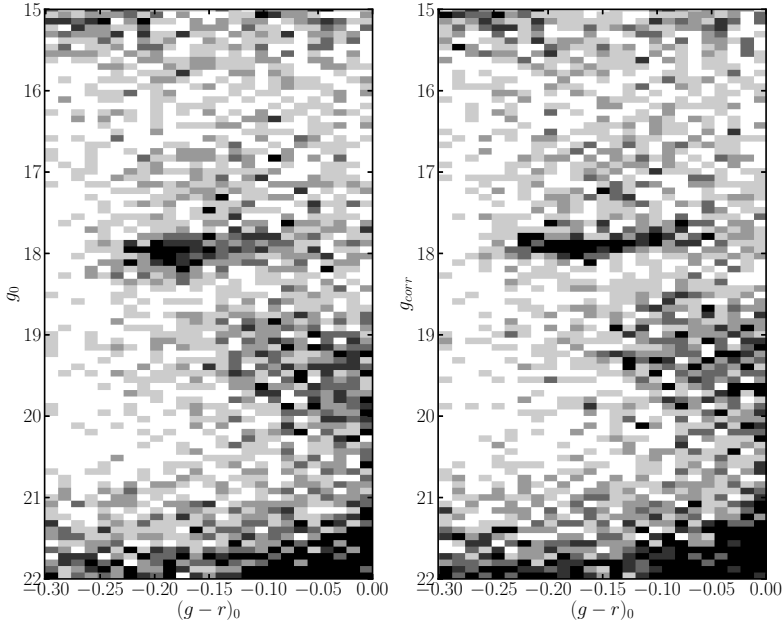}
\caption{{\it Left panel}: Hess diagram depicting the initial photometric
selection of BHB stars: $120\arcdeg < l < 165\arcdeg$, $b <
-40\arcdeg$, $0.8 < (u-g)_0 < 1.5$, and $-0.3 < (g-r)_0 < 0$ with a
color-color selection from \citet{ynk+00} to separate BHBs from BS
stars.  Darker shading indicates higher density of stars. The selection was
centered around $l \sim 143\arcdeg$ to contain the spatial location
of CPS debris as found by \citet{nyw09}.  A population of likely BS stars is evident at redder
colors ($(g-r)_0 > -0.15$) at fainter magnitudes ($g_0 \sim 19-20$) than the
main BHB population. These BS stars are likely associated with the Sgr
stream, as was noted by \citet{kbe+12}. {\it Right panel:} Hess
diagram of the same stars seen in the left panel, but with magnitudes
corrected by Equation~\ref{gcorr_eqn} for the distance modulus trend
with longitude, as discussed in the text. The BHB is much narrower in
this panel, with little dependence of magnitude on the color.}
\label{hpb}
\end{figure}

We take advantage of this relationship between latitude and distance for the
CPS to study
the properties of the BHB stars in the stream. The left panel of
Figure~\ref{hpb} shows the observed Hess diagram for blue ($-0.3 <
(g-r)_0 < 0.0$) stars in the vicinity of the CPS (specifically,
$120\arcdeg < l < 165\arcdeg, b < -40\arcdeg$), with an additional
color-color selection from \citet{ynk+00} to more prominently select
BHB stars from among the more numerous blue stragglers. The BHB of
Cetus stars is clearly visible at $g_0 \sim 18$, with an additional
large blob of stars at redder ($-0.15 \lesssim (g-r)_0 < 0.0$) colors and
fainter ($g_0 \sim 19-20$) magnitudes. As noted by \citet{kbe+12}, this
feature is likely made up of BS stars from the Sgr dSph tidal stream,
which intersects the CPS in this region of the sky. As shown in the
previous paragraph and Figure~\ref{gmodbpb}, the latitude dependence
of CPS BHB stars' magnitudes can be eliminated. The right panel of
Figure~\ref{hpb} shows the same stars as the left panel, but with
corrected magnitudes, $g_{corr}$, instead of the measured magnitudes.
The BHB locus in the ``corrected'' panel is noticeably narrower than
the original, and the absolute magnitude of BHB stars in the CPS appears to be only weakly color-dependent. The BS feature is relatively
unchanged, or perhaps even more diffuse, as would be expected to
happen for a population with a different distance distribution than
the CPS when applying the distance modulus correction.

\section{The position of the CPS in the sky}\label{sec:position}

\begin{figure}[!t]
\plotone{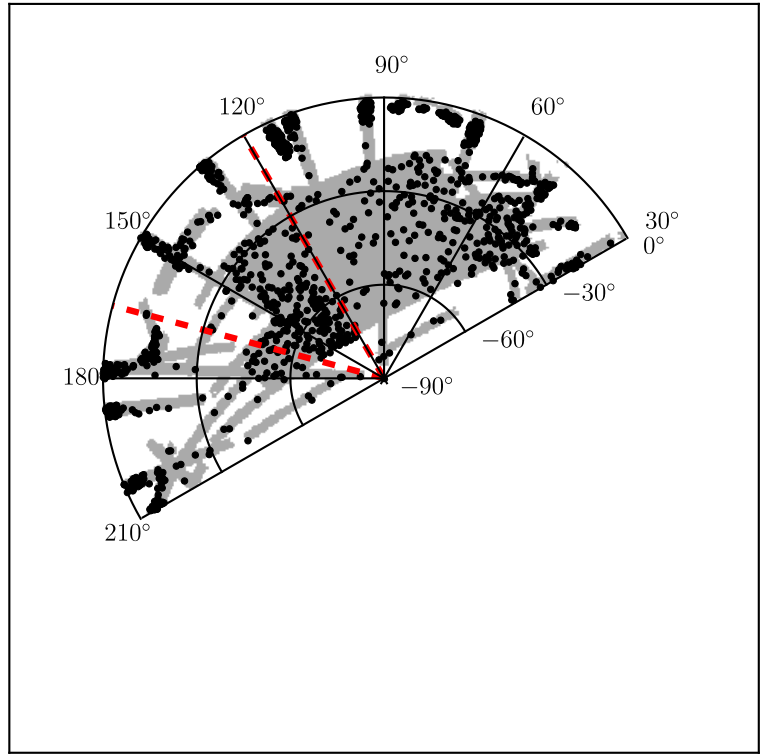}
\caption{Galactic coordinates of color-selected BHB stars (black dots) that are
within 0.1 magnitudes of the quadratic relationship in $D_{mod}$
vs. $b$, as shown in Figure~\ref{gbpb} and Equation~\ref{dm_eqn},
are shown in a polar plot of the south Galactic cap.  Dashed lines
show the region in $l, b$ space from which the BHB stars in
Figure~\ref{hpb} were selected.  The southern footprint of the SDSS
DR8 photometric catalog is plotted in the background in grey. It is
clear that BHB stars cluster around a region centered at $l \sim 143
\arcdeg$, in agreement with \citet{nyw09}, though the incomplete
coverage of DR8 for $l > 150\arcdeg$ makes it difficult to assess
whether the CPS extends to higher Galactic longitudes.  Again, the CPS
appears to drop off rapidly at latitudes above $b=-40\arcdeg$.}
\label{lbppb}
\end{figure}

We now examine the positions of photometrically-selected BHB stars to
trace the path of the Cetus stream on the sky. In Figure~\ref{lbppb}
we show a polar plot in Galactic coordinates, centered on the south
Galactic cap, of the sky positions of BHB stars selected within 0.1
magnitudes of the trendline we fit to the concentration of BHBs in
Figure~\ref{gbpb} (except we have removed the Galactic longitude constraint). There is a clear concentration of stars between $l
\sim 120\arcdeg$ and $\sim 160\arcdeg$ relative to the number of BHBs
in adjacent longitude regions. In \citet{nyw09} the CPS was found to
have a nearly constant longitude of $l \sim 143\arcdeg$ along its
entire length. We aim to use the additional information that is
available in this region from DR8 to revise the positional
measurements of the CPS. However, two complications are obvious in
Figure~\ref{lbppb}: first, the Sagittarius stream intersects the CPS
at $b \sim -70\arcdeg$ at nearly the same distance as Cetus, and
secondly, the DR8 footprint cuts off at $l \sim 150\arcdeg$, making it
difficult to determine whether the CPS extends beyond this longitude.

\begin{figure}[!t]
\plotone{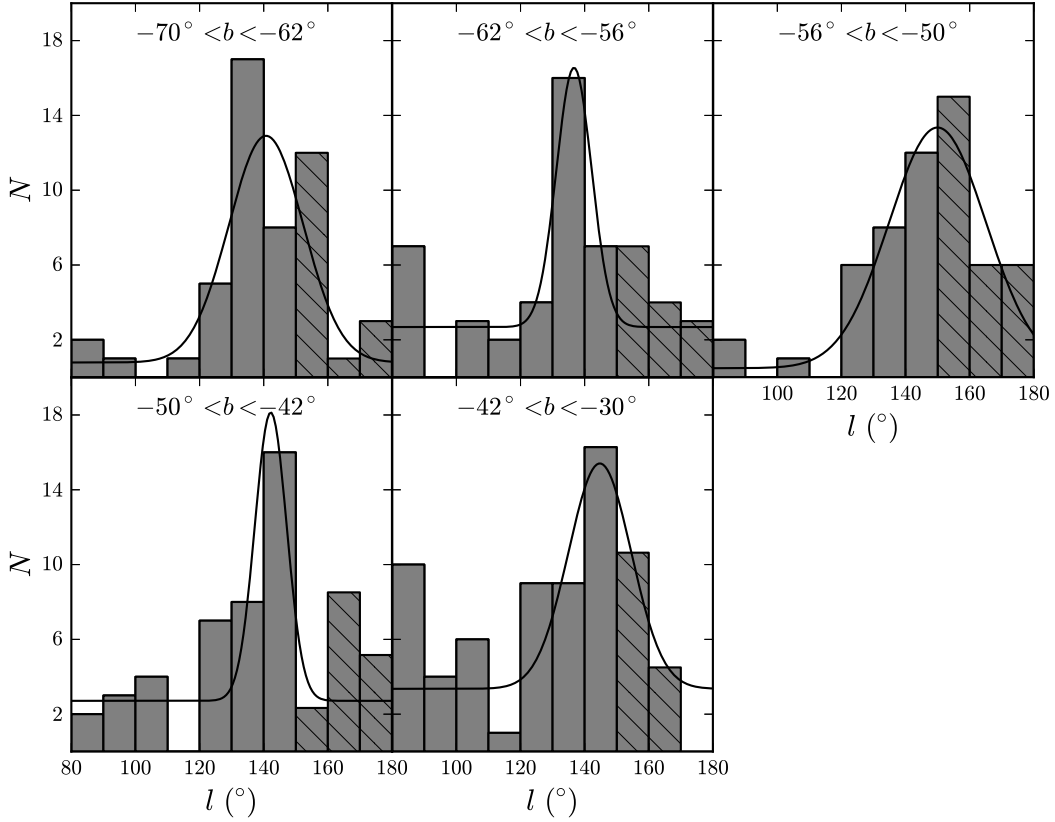}
\caption{The color and distance modulus-selected BHB stars from
Figure~\ref{lbppb} were divided into the following latitude bins:
$-70\arcdeg<b<-62\arcdeg$, $-62\arcdeg<b<-56\arcdeg$,
$-56\arcdeg<b<-50\arcdeg$, $-50\arcdeg<b<-42\arcdeg$, and
$-42\arcdeg<b<-30\arcdeg$.  Number counts of BHB stars in each
latitude strip were calculated in $10\arcdeg$ bins of longitude
between $80\arcdeg<l<180\arcdeg$.  Bin heights (N) for incomplete bins
were scaled upward by the corresponding fraction of area covered by
DR8 photometry such that the new bin height is N/(Fractional Area). Bins that have been corrected for incompleteness are denoted with diagonal crosshatching.}
\label{blh}
\end{figure}

\begin{figure}[!t]
\plotone{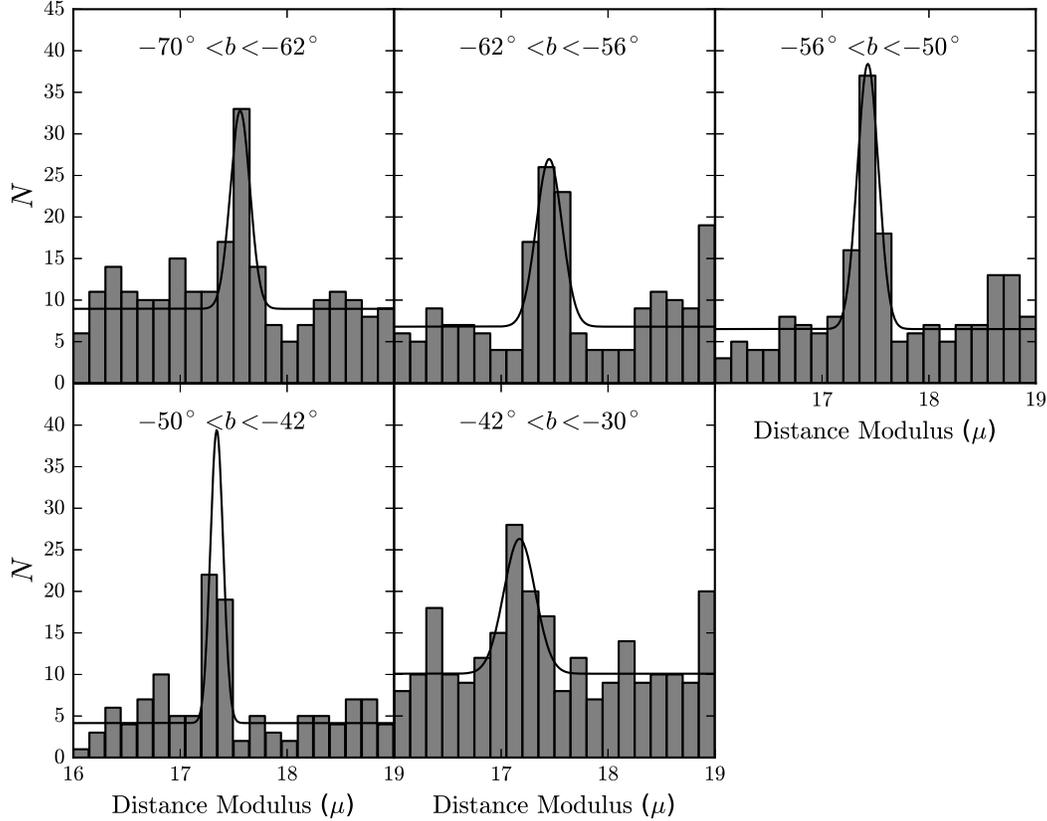}
\caption{Distance modulus of BHB stars between $120\arcdeg < l < 165\arcdeg$ in five latitude bins. A narrow peak is seen in each latitude bin that likely corresponds to CPS stars. A Gaussian fit to the peak is shown in each panel; the results of these fits are given in Table~\ref{cpsk}, and provide the distances used throughout this work.
}
\label{fig:dmod}
\end{figure}

To measure the position of the CPS, we slice the BHB sample into five
latitude bands, then plot histograms of the longitude distribution in
each strip. These histograms are seen in Figure~\ref{blh} for BHB
candidates within 0.1 magnitudes of the distance modulus trend fit
previously (Equation~\ref{dm_eqn}). The latitude ranges were selected
to include roughly the same number of stars in each range, with strips
centered on $b = [-66\arcdeg, -59\arcdeg, -53\arcdeg, -46\arcdeg,
-36\arcdeg]$.  Bins with incomplete photometric coverage in DR8 were
corrected upward by dividing the bin height by the fractional sky area
covered by the bin in SDSS, with uncertainties in the bin heights
corrected appropriately.  A peak is evident in each panel of the
figure, which we attribute to the CPS.  We fit a Gaussian to the peak
in each panel.  The fitting was performed using a standard Gaussian
function with an additional constant offset in $N$ as a fit
parameter. This parameter represents the unknown background level
present in the data. A simple, coarse grid-search was performed over
the expected parameter ranges to determine the general location of the
global best-fit of each dataset; a gradient descent search was then
started near the global best-fit parameters (as indicated by the grid
search). The final parameters are those determined by the gradient
descent, with the model errors at those parameters given by the
square-roots of twice the diagonal elements of an inverted Hessian
matrix.  We find best-fit central Galactic longitudes of $l =
[140.7\pm3.5\arcdeg, 136.7\pm4.9\arcdeg, 150.0\pm4.6\arcdeg,
142.2\pm4.9\arcdeg, 144.7\pm4.7\arcdeg]$, respectively, for each
of the latitude strips. In Table~\ref{cpsk} we give the best fit positions with their associated errors as $l$. We note that the uncertainties in these
positions are larger than those given by \citet {nyw09}, but
consistent within the error bars. The width of the stream (in degrees) is the 
best-fitting Gaussian $\sigma$; these values are tabulated as $\sigma_l$ in Table~\ref{cpsk} along with the positions and other stream parameters. Because the BHB stars are rather sparse tracers of the stream, the stream widths are easily biased by background fluctuations toward large ($5\arcdeg - 15\arcdeg$) values that are likely overestimates of the true stream width. 

Distances to the CPS at each of these measured positions were
determined in a similar manner to the positions. In each of the five latitude bins, we selected all 
BHB candidates between $120\arcdeg < l < 165\arcdeg$. In each latitude bin, the distance moduli 
$D_{\rm mod}$ (calculated using Equation~\ref{dm_eqn}) of the BHB stars were histogrammed, 
and we fit a Gaussian to their distribution using the gradient descent method described above. We 
restricted the fits to $16 < D_{\rm mod} < 19$ in each bin, within which a peak was clearly visible in 
all five ranges considered. These histograms and the Gaussians fit to the peaks in each latitude bin are shown in Figure~\ref{fig:dmod}. The distance moduli and their Gaussian spread are given as $D_{\rm 
mod, fit}$ and $\sigma_{D_{\rm mod, fit}}$ in Table~\ref{cpsk}, with the associated distances and line-of-sight depths reported as d$_{\rm fit}$ and $\sigma_{\rm d, fit}$. These are likely much more
robust measurements of the stream's physical width than the spatial
distributions on the sky ($\sigma_l$ and the associated width $\sigma$ in kpc), as the distinct peaks in distance modulus are much less prone to
contamination by non-stream stars than the positions on the sky. The line-of-sight depths are
narrower than the widths of the stream on the sky in all cases.
The stream width will be discussed further in 
Section~\ref{sec:nbody}, where we compare the measurements to N-body model results.

It might seem obvious, once the position and distance to CPS are well
known, to look for the much more numerous F turnoff stars associated
with it.  We did make some unsuccessful attempts to do this.  The F
turnoff stars are expected to be 3.5 magnitudes fainter than the
horizontal branch, at $g_{corr} \sim 21.5$.  The difficulty is that
there is a very large background of F turnoff stars in the halo and
particularly in the Sgr dwarf tidal stream, which dominates the sky
when we attempt to extract F turnoff stars.  We were successful in
tracing the CPS in BHB stars because there are relatively few BHB
stars in the Sgr stream compared to CPS.  This is apparently not the
case for F turnoff stars.

\section{The Metallicity of the CPS}\label{sec:feh}

We now use the CPS distance estimates and the velocities from
\citet{nyw09}, to select luminous stars from the Cetus Polar Stream.
From these stars, we will study the range of metallicities in the
stream.

\begin{figure}[!t]
\includegraphics[width=2.6in]{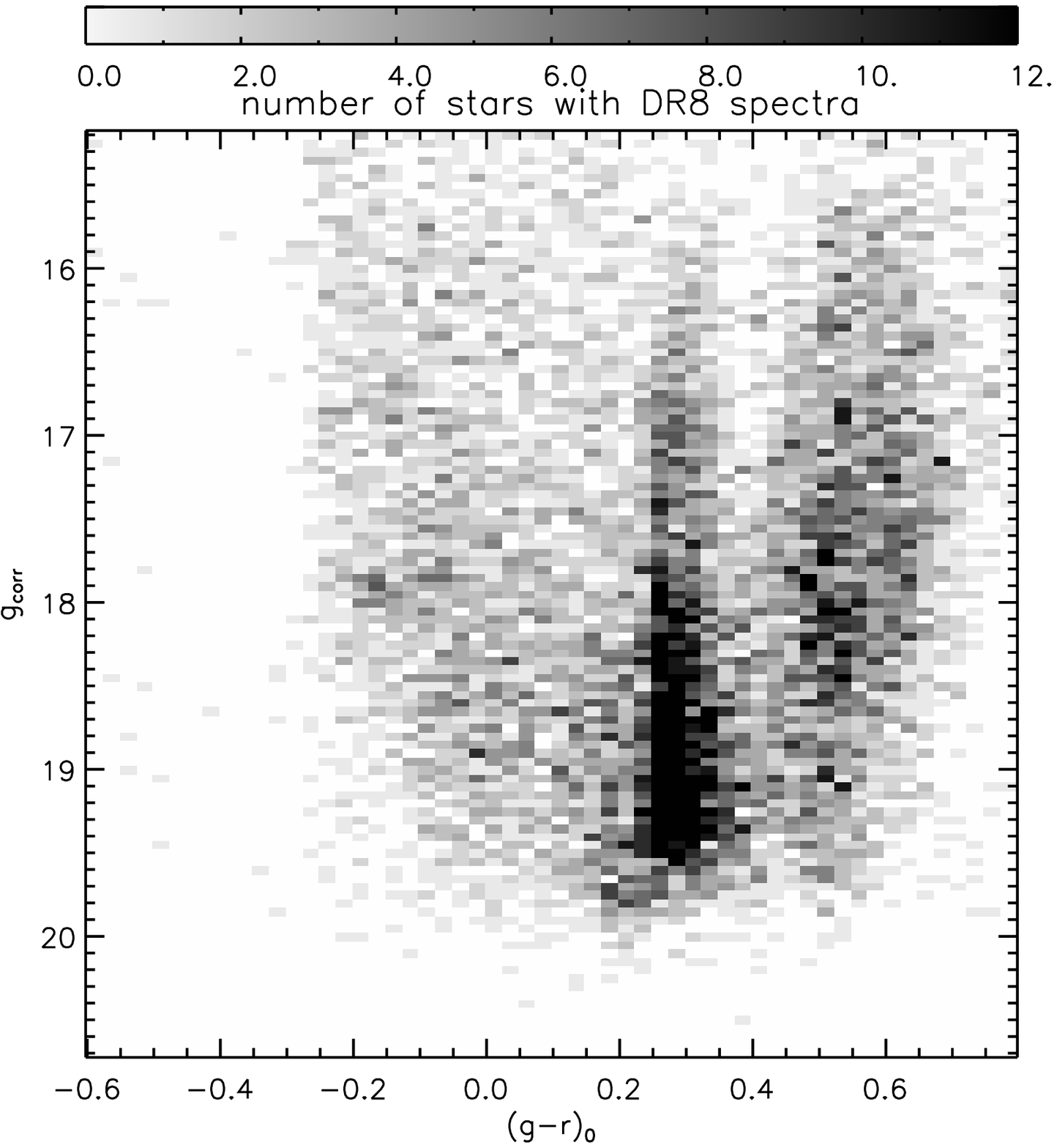}
\includegraphics[width=4.0in]{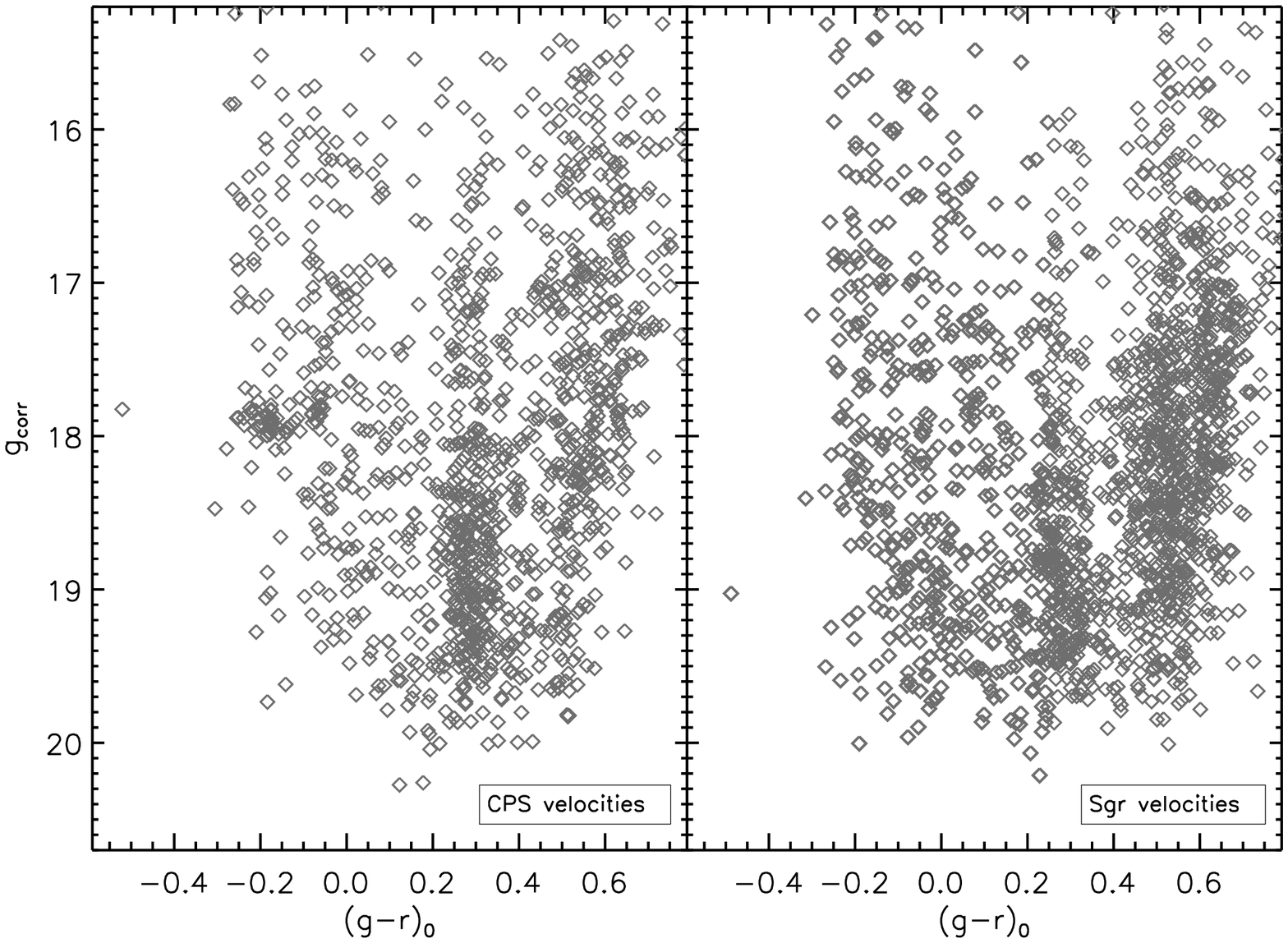}
\caption{{\it Left panel:} Hess diagram of all stars having SDSS DR8 spectra in the south Galactic cap ($b<0\arcdeg$) that have surface
gravity measurements consistent with classification as giant stars
($1.0<\log{g}<4.0$), proper motions consistent with zero ($\left|
\mu_b \right| < 6$ mas yr$^{-1}$, $\left| \mu_l \right| < 6$ mas
yr$^{-1}$; to select against nearby stars with large tangential
motions), and have low metallicity ($-4.0<$[Fe/H]$<-1.0$).  
The apparent magnitudes in this CMD have been latitude corrected, as defined in
Figure~\ref{gmodbpb} and Equation~\ref{gcorr_eqn}.
{\it Center panel:} Stars with the same selection criteria as in the left panel that also have velocities consistent with the CPS (selected using the CPS velocity-selection criteria from
\citealt{nyw09}). The BHB, RGB, and possibly the asymptotic giant branch of the CPS stand out
among stars with Cetus-like velocities in this diagram.
{\it Right panel:} As in the center panel, but for stars with Sgr tidal stream velocities (using selection criteria from \citealt{nyw09}). Because apparent
magnitudes in this CMD have been latitude corrected for the distance
to the CPS (not Sgr), we do not necessarily expect the Sgr stars to
form clear sequences in this diagram.  However, we can see that there
are generally a large number of blue straggler and giant branch stars
at approximately the right apparent magnitudes, and there are a few
BHB stars that are likely members of the Sgr trailing tidal tail.
}

\label{csi}
\end{figure}

We select stars with spectra in SDSS DR8 that are in the south
Galactic cap.  With a horizontal branch at $g_0=18$, the turnoff of
the CPS is at approximately $g_0=21.5$.  Since this is fainter than
the spectroscopic limit of SDSS, we do not expect to find any main
sequence CPS stars in the dataset.  Therefore, we select only luminous
giant stars, using the surface gravity criterion $1.0 < \log {g} <
4.0$.  Here, $\log{g}$ was determined from the ELODIELOGG value in the
SDSS database.  At distances of about 30 kpc, we do not expect CPS
stars to have a significant tangential velocity, so we also used the
proper motion cut $\left| \mu \right| < 6$ mas yr$^{-1}$, which
selects SDSS objects whose proper motions are consistent with zero.
We also eliminated nearby Milky Way stars with high metallicity by
selecting only those with $-4.0<$ [Fe/H] $<-1.0$.  The metallicity of
the sample was calculated differently for stars with $(g-r)_0<0.25$
and $(g-r)_0>0.25$.  This was because \citet{nyw09} find the WBG
classification (FEHWBG; \citealt{wbg99}) to be a better measure of metallicity for BHB
stars, and thus the WBG metallicity was used for $(g-r)_0<0.25$ while
the adopted SDSS metallicity (FEHA; consisting of a combination of a
number of different measurement techniques) was used for
$(g-r)_0>0.25$. The DR8 data we downloaded contained multiple entries for some stars. For these stars, we combined multiple measurements by calculating a weighted mean velocity (weighted by the SDSS velocity errors). Their stellar parameters ([Fe/H] and $\log {g}$) were set to the values of the highest $S/N$ measurement.

The left panel of Figure~\ref{csi} shows the latitude-corrected Hess diagram of
$\log{g}$-selected metal-poor giant star candidates in the south Galactic
cap. The greyscale represents the density of stars satisfying the surface
gravity, metallicity, and proper motion criteria. In the center panel, we show a CMD of those stars from the left panel that also have the expected velocity of the CPS.
To select stars with CPS-like velocities, we used $V_{gsr} < (-0.1818
\times b - 33.63)$ km s$^{-1}$ and $V_{gsr} > (-1.205 \times b -
130.36)$ km s$^{-1}$ and $V_{gsr} > (2.91 \times b + 13.67)$ km
s$^{-1}$, where the constraints are those used by \citet{nyw09} to
select Cetus members (note: here, and throughout this work, the
subscript ``gsr'' means that the velocities are along the line-of-sight, and relative to the
Galactic standard of rest).  One sees among the velocity-selected CPS
candidates in this figure a strikingly narrow BHB (centered around
$(g-r)_0 \sim -0.15$), red giant branch (at $0.45 \lesssim (g-r)_0 \lesssim
0.7$ and $16.5
\lesssim g_{corr} \lesssim 19$), and what appears to be an asymptotic giant branch at $g_{corr} 
\sim 17$ and $(g-r)_0 \sim 0.5$.

\begin{figure}[!t]
\includegraphics[width=4.25in]{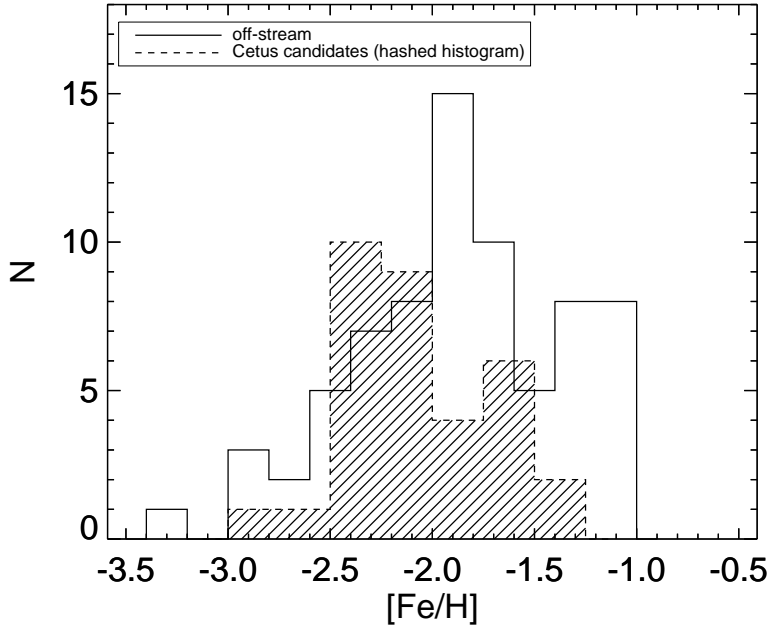}
\caption{Metallicity distribution of  BHB candidates selected with
colors of $-0.25 < (g-r)_0 < -0.05$, low surface gravity, and low proper motions. The hashed (dashed-line) histogram shows stars that are likely members of the CPS, selected within 0.1 magnitudes of the
trendline defined in Figure~\ref{gbpb}, with CPS velocities, and with
$120\arcdeg<l<165\arcdeg$.  A narrow peak is visible between $-2.5 \lesssim {\rm [Fe/H]} \lesssim -2.0$. The open (solid-line) histogram represents stars outside the spatial and velocity cuts of CPS stars, which provide a background comparison. The Cetus candidates have a lower mean metallicity than typical field BHBs in this area of the sky.} 
\label{fh}
\end{figure}

To assess the possible contamination of the CPS sample by Sgr stars,
we plot a similar color-magnitude diagram in the right panel of Figure~\ref{csi}. The points in this CMD were selected from the box
chosen by \citet{nyw09} to highlight stars with Sgr velocities. Sgr
velocities were chosen using $V_{gsr} < (-3.01 \times b - 271)$ km
s$^{-1}$ and $V_{gsr} > (-4.12 \times b - 395.25)$ km s$^{-1}$ for $b
< -52\arcdeg$ and $-180 < V_{gsr} < -114.48$ km s$^{-1}$ for $b >
-52\arcdeg$.  Neither a clear BHB, nor any other obvious feature, is
visible in this figure. There are possibly excess red
giants among the velocity-selected Sgr candidates, but they do not
form a clear sequence in the CMD.  This is as expected if Sgr is not
at the same distance as the CPS, since we have shifted things to CPS
distances by the use of $g_{corr}$; this should smear out any Sgr
features that are present in such a CMD.

\begin{figure}[!t]
\plotone{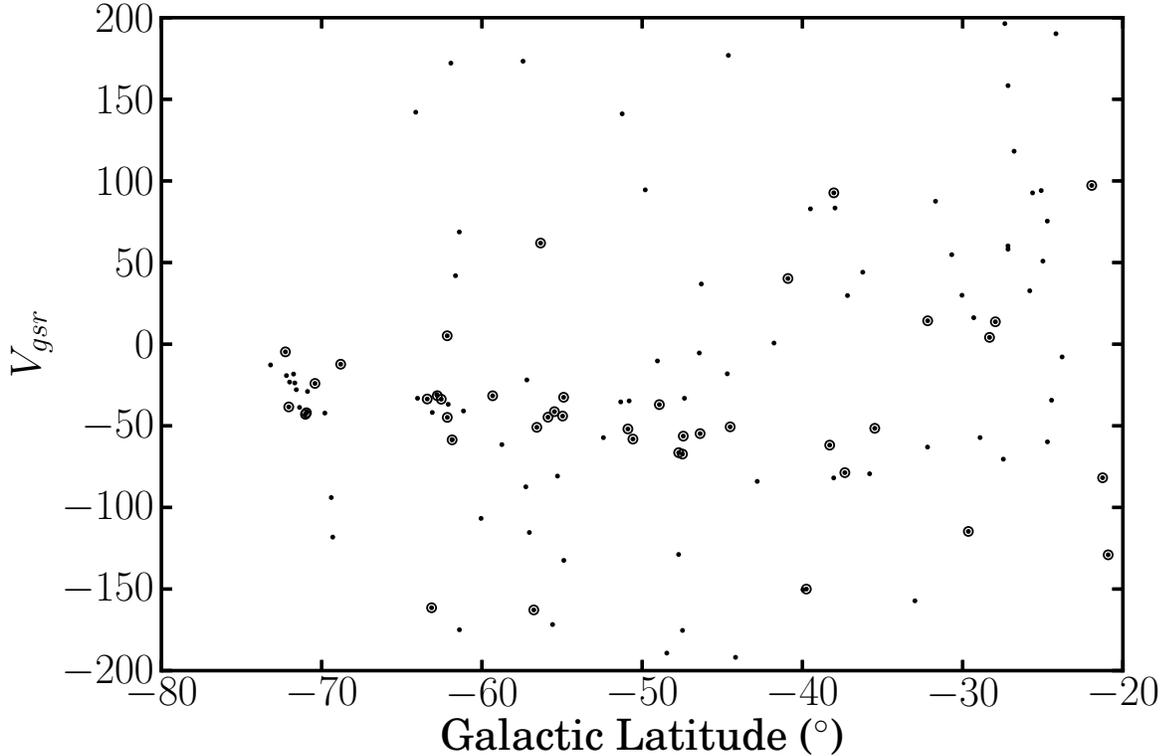}
\caption{Illustration of the effectiveness of selecting CPS BHB stars by metallicity.
Dots represent BHB stars selected by surface gravity and low
proper motion, whose distance is within 0.1 magnitudes of the
(Galactic latitude-dependent) distance modulus of the CPS.  Larger circles highlight stars with metallicities
consistent with the CPS ($-2.5 < $[Fe/H]$_{WBG} < -2.0$). A trend is
visible that stretches from $V_{gsr} \sim -30$~km~s$^{-1}$ at $b = -70\arcdeg$ to $V_{gsr} \sim -65$~km~s$^{-1}$ at $b = -45\arcdeg$ and perhaps beyond. The
handful of stars at $V_{gsr} < -100$~km~s$^{-1}$ correspond to Sgr
velocities.}
\label{vvbsb}
\end{figure}

In Figure \ref{fh}, we show a histogram of metallicities of the BHB stars from
Figure~\ref{csi}. The hashed histogram is made up of CPS candidates that have $-0.25<(g-r) _0<-0.05$, distance moduli
within 0.1 magnitude of the CPS trendline, $V_{gsr}$ within 20 km~s$^{-1}$ of the velocity trend used to select candidates in Figure~\ref{csi}, and are between $120\arcdeg<l<165\arcdeg$. The open (solid line) histogram is a background sample, selected from outside both the CPS spatial region and velocity criteria. The CPS BHB stars occupy a narrow metallicity range, with a peak around [Fe/H]$=-2.2$, that is clearly unlike the metallicity distribution of the background sample (typical uncertainties on the metallicity for each star are $\sim0.25$ dex).
Thus, later in this paper, we will use $-2.5<$ [Fe/H] $<-2.0$ to
preferentially select stars that are in the CPS.

To illustrate the effect of metallicity selection, we show in
Figure~\ref{vvbsb} the gsr-frame line-of-sight velocity as a function of Galactic latitude for BHB stars
(selected by low surface gravity and low proper motion) that have
$-0.25<(g-r)_0<-0.05$ and are within 0.1 magnitudes of the correct
distance modulus to be members of the CPS.  We did not select the
stars based on Galactic longitude.  The larger dots in this figure are
the stars with $-2.5<$ [Fe/H] $<-2.0$.  Note that the CPS stands out
much more clearly in the metallicity-selected sample, though we
probably lose a few bonafide CPS stars.  Using the low metallicity
sample and our knowledge of the distance to the CPS, we have very
little contamination from Sgr and other BHB stars in the Milky Way,
even though we accept stars from all observed Galactic longitudes.

\section{Red Giant Branch Stars in the CPS}\label{sec:rgb}

In the center panel of Figure~\ref{csi} there are apprently many Red Giant
Branch (RGB) stars in the CPS, that we would like to add to our
sample.  Using the knowledge that CPS stars are metal-poor ($-2.5 <
$[Fe/H]$ < -2.0$), we choose to select candidates using a fiducial sequence from the
globular cluster NGC 5466, which has a metallicity ([Fe/H]) of -2.22,
shifted to a distance modulus of 17.39 (corresponding to a distance of
30.06 kpc, which is the distance at $b = -50\arcdeg$ from the fit in
Figure~\ref{hpb}).  The fiducial sequence of NGC 5466 is from SDSS
data, and is taken from \citet{ajc+08}.  A third order polynomial was
fitted to the region of interest on the fiducial sequence of NGC 5466,
and stars were selected using the criteria of $-0.0081006g_{corr}^3 +
0.46944g_{corr}^2 -9.1207g_{corr} + 59.9130 -0.04 < (g - r)_0 <
-0.0081006g_{corr}^3 + 0.46944g_{corr}^2 -9.1207g_{corr} + 59.9130 +
0.04$ and $16.9 < g_{corr} < 19.9$, where $g_{corr}$ is the latitude
corrected magnitude as defined in Equation~\ref{gcorr_eqn}.

\begin{figure}[!t]
\includegraphics[height=5.0in]{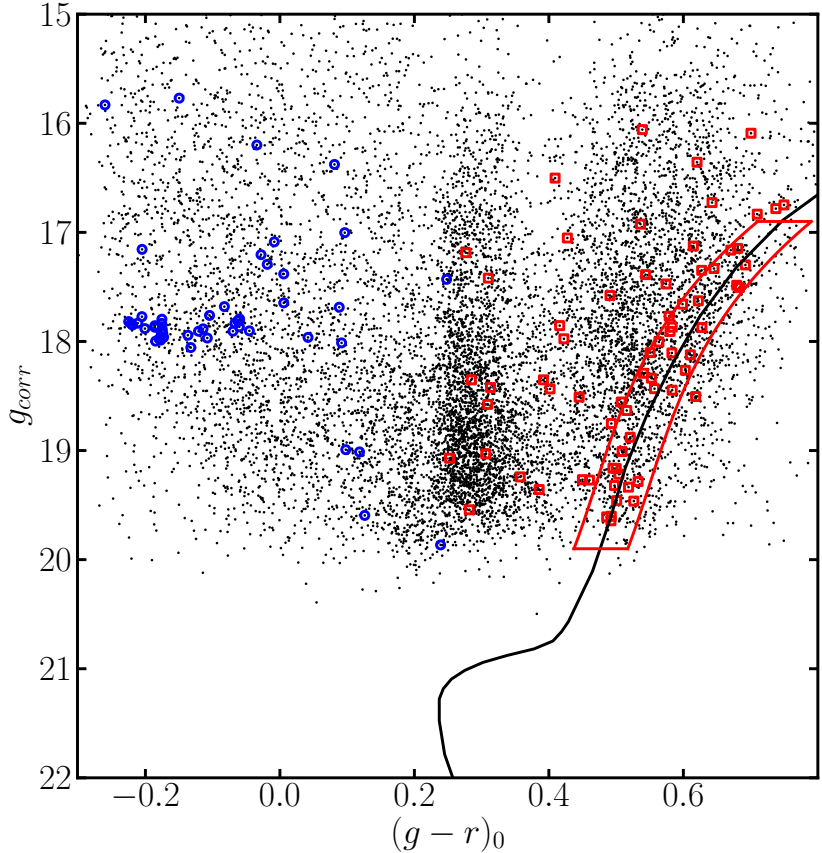}
\caption{Color-magnitude diagram of stars in the south Galactic cap region
(black points) with low surface gravity, low proper motions consistent
with their being Milky Way halo giants, and low metallicity ($-4.0<$
[Fe/H] $<-1.0$). The colored (blue and red) circles represent low
metallicity stars likely to be CPS members. These stars have $-2.5 <
$[Fe/H]$ < -2.0$, velocities within 20 km s$^{-1}$ of the final fit
of $V_{gsr}$ as a function of latitude for the CPS (which we show
later in the paper, in Figure~\ref{vvbsf2}), and are within longitudes
of $120\arcdeg < l < 165\arcdeg$.}
\label{csf2}
\end{figure}

Figure~\ref{csf2} shows a CMD similar to those in
Figure~\ref{csi} for spectroscopically selected giant
stars between $120\arcdeg < l < 165\arcdeg$ with low proper
motion. Colored points show those with metallicities of $-2.5<$ [Fe/H]
$<-2.0$ and CPS velocities. We note that the velocity selection shown
here for CPS is actually the final selection we arrive at later in
this paper; however, the final result differs little from the previous
measurement of the velocity trend by \citet{nyw09}.  The
color-magnitude box described above for selecting CPS RGB stars, along
with the NGC~5466 ridgeline upon which it is based, is shown on the
diagram.

To determine the metallicity of the CPS, we used BHB stars, which are
by their nature all low metallicity.  We then selected red giant
branch stars with the same metallicity range as the BHBs before we
matched the RGB fiducial.  It is natural to wonder, then, whether
there are any higher metallicity RGB stars in the CPS.

\begin{figure}[!t]
\includegraphics[width=6.75in, trim= 1.0in 0.0in 1.0in 0.25in, clip]{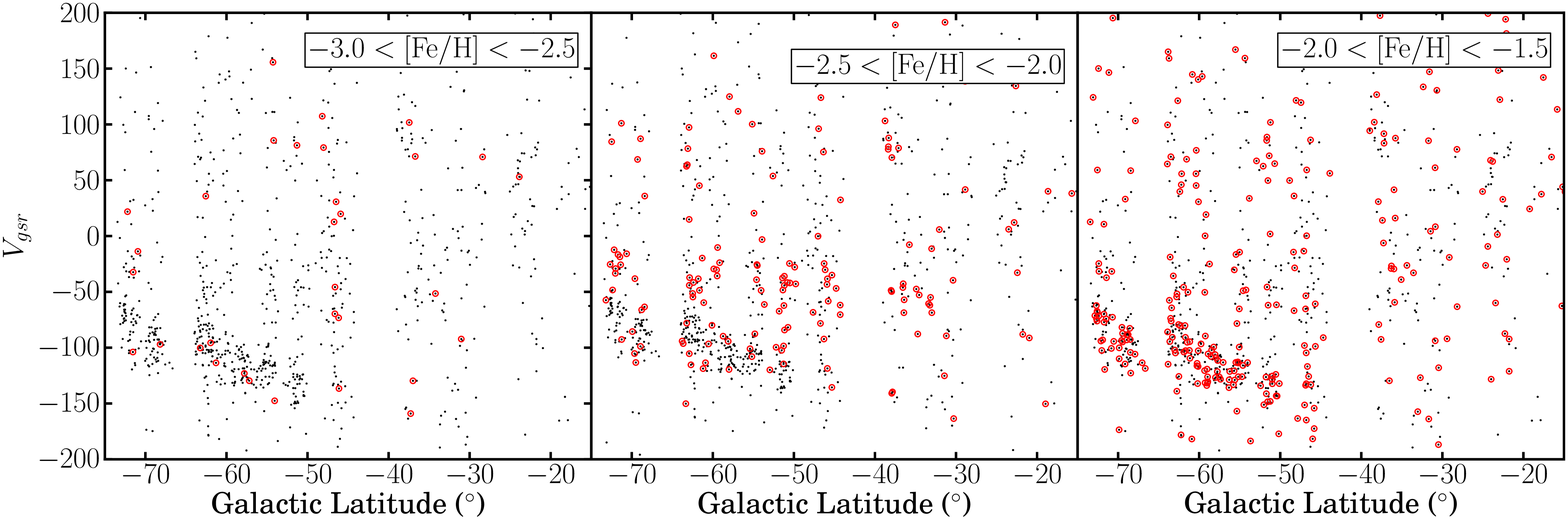}
\caption{To show that the metallicity range of $-2.5 <$ [Fe/H] $< -2.0$
captures the most significant portion of the CPS, three graphs of
$V_{gsr}$ vs. $b$ for RGB stars are shown here highlighting different
metallicity ranges. These stars are all selected between $0.4 <
(g-r)_0 < 0.8$ and $16.9 < g_{corr} < 19.9$ to encompass giant branch
stars at a large range of distances. Black points represent all stars
within these ranges satisfying all of the CMD and spatial cuts as used
before, while large red circles in each panel show stars with $-3.0 <$
[Fe/H] $< -2.5$, $-2.5 <$ [Fe/H] $< -2.0$, and $-2.0 <$ [Fe/H] $<
-1.5$, respectively, from left to right. The left panel shows very few
stars with [Fe/H]$<-2.5$. In the middle panel, the metallicity range
we found (via Figure~\ref{fh}) to best represent CPS BHB stars seems
to select predominantly CPS stars following the stream velocity trend,
with very few Sgr stars in the $-2.5 < $[Fe/H]$ < -2.0$ range. The
right panel, in contrast, shows that most Sgr RGB stars are at higher
metallicity than those in the CPS. Clearly, the initial metallicity
selection brought forth by the BHB data highlights the strongest
signal of a stream in the expected region.}
\label{vvbsm}
\end{figure}

In Figure~\ref{vvbsm}, we show the line-of-sight, Galactic standard of
rest velocities as a function of Galactic latitude for all of the low
surface gravity, low proper motion stars with spectra in SDSS DR8,
that are within the Galactic longitude limits
$120\arcdeg<l<165\arcdeg$ and within the box $0.4<(g-r)_0<0.8$ and
$16.9<g_{corr}<19.9$. The three panels of this figure show all stars
satisfying these criteria as black points, with large red dots
representing stars with metallicities between $-3.0 <$ [Fe/H] $<
-2.5$, $-2.5 <$ [Fe/H] $< -2.0$, and $-2.0 <$ [Fe/H] $< -1.5$ in
panels from left to right, respectively. Notice that we have opened up
the range of colors and apparent magnitudes very wide, to accept all
types of giant branch stars at a range of distances.

We see from this figure that the much more populated Sgr dwarf tidal
stream, in the lower left corner of each panel, includes stars of many
different metallicities, and is most pronounced in the metallicity
range $-2.0<$ [Fe/H] $<-1.5$.  On the other hand, the CPS, with
velocities near $V_{gsr}\sim -50$ km s$^{-1}$, is most apparent in the
center panel, with $-2.5<$ [Fe/H] $<-2.0$ (with perhaps a few slightly
more metal-rich stars).  The population of stars we see in the CPS
includes predominantly metal-poor stars in a narrow range of metallicity.  We
also tried selecting stars with CPS velocities in color-magnitude
fiducial sequences with a range of metallicities, but the stars in
each fiducial sequence were dominated by stars in the same metallicity
range of $-2.5 <$ [Fe/H] $<-2.0$.  Therefore our initial fiducial
sequence of NGC~5466 as well as the selections in the CMD and in
metallicity are justified.

If the CPS is the remnant of a dwarf galaxy, then either we are
looking at the outer portions of the dwarf galaxy, and not the inner
portion with more recent star formation, or we are looking at the
remains of a smaller, possibly gas-stripped galaxy that never had a
later epoch of star formation at all (similar to the ultra-faint
dwarfs discovered recently that seem to have had only a single epoch
of star formation).

\section{The line-of-sight velocities along the CPS}

\begin{figure}[!t]
\plotone{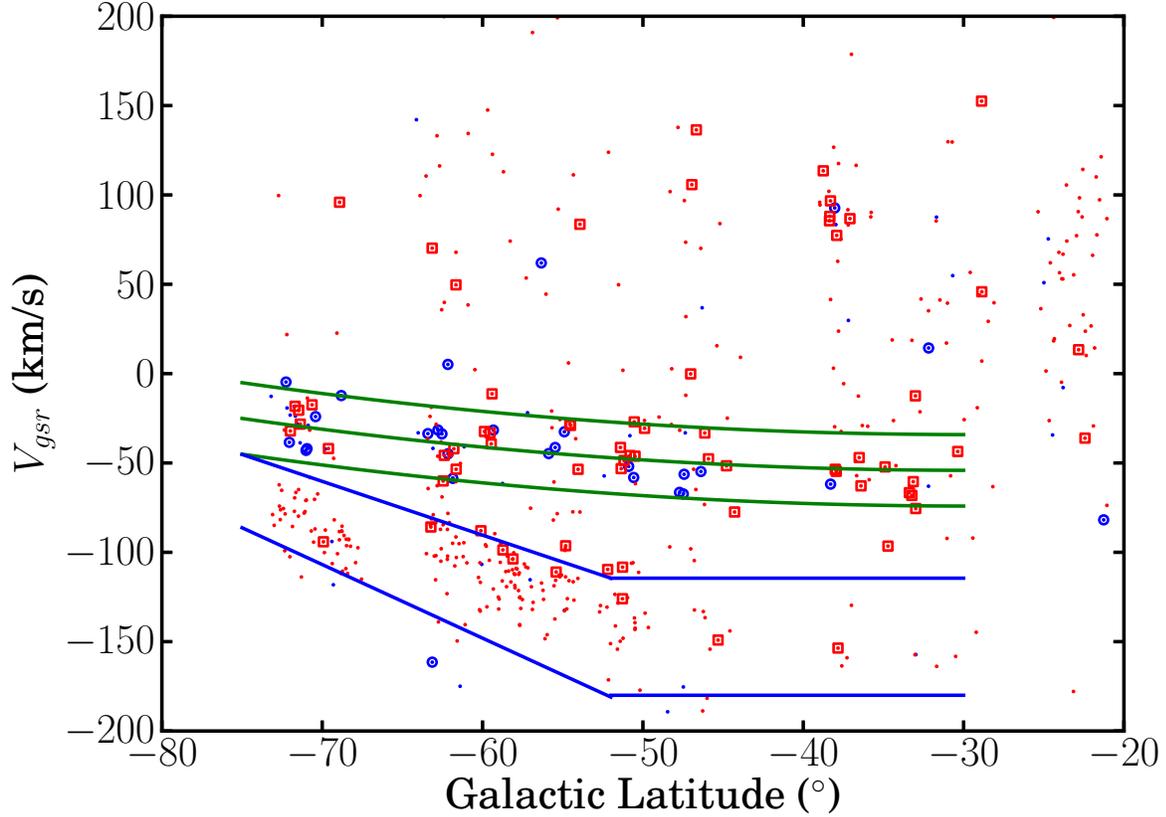}
\caption{Galactic standard of rest velocities of giant stars (shown as colored points in Figure~\ref{csf2}) as a function of Galactic
latitude. These stars are selected to have low surface gravity and low
proper motion, with metallicities in the range of $-4.0 <$ [Fe/H]
$<-1.0$, and lie in the region $b < 0\arcdeg$ and $120\arcdeg < l <
165\arcdeg$. Blue and red points represent BHB and RGB candidates, respectively. Larger symbols are stars with metallicities representative
of CPS stars ($-2.5 <$ [Fe/H] $< -2.0$); BHBs are large blue circles, and RGB: red squares.
The central solid green line represents the polynomial that
we fit to the highlighted stars: $V_{gsr} = -41.67 - (0.84 \times b) - (0.014 \times b^2)$.  Upper and lower bounds (also
shown as solid green lines) of 20 km~s$^{-1}$ on either side of this
trendline were used to select CPS stars by their $V_{gsr}$. The region
enclosed within the blue lines represents the selection of Sgr
velocities, as discussed in \citet{nyw09}.  The clump of stars at $b=-38^\circ$ and $V_{gsr}=95$~km~s$^{-1}$ belong to the Pisces Stellar Stream \citep{bgk12, mcn+13}.}
\label{vvbsf2}
\end{figure}

We have now identified samples of BHB and RGB stars in the CPS that lie within the appropriate
loci in Figure~\ref{csf2}.  In Figure~\ref {vvbsf2} we show the
line-of-sight, Galactic standard of rest velocities for these stars
(removing the selection in $V_{gsr}$) as a function of Galactic
latitude.  To review, the stars in this plot are low surface gravity,
low proper motion stars with spectroscopy in the SDSS DR8. They
additionally are in the region of the sky inhabited by the CPS
($b<0\arcdeg$ and $120\arcdeg<l<165\arcdeg$).  The colors of the
symbols tell which CMD selection box each star is from: blue points 
are BHB candidates, and red: RGB. The larger symbols
in the figure have the metallicity we have shown to preferentially select CPS stars:
$-2.5<$[Fe/H]$<-2.0$; large blue circles are BHB stars, and red squares are RGB candidates.  The majority of the points in
Figure~\ref{vvbsf2} that have larger-sized symbols follow the expected
velocity trend of the CPS, with only a handful of stars near the Sgr
velocities, and a few scattered elsewhere.

We fit a polynomial to the CPS velocities beginning with the entire
metallicity-selected ($-2.5 < $ [Fe/H]$ < -2.0$) sample. The fitting
was done iteratively, rejecting outliers at each iteration using the
same technique as in Section~2, until it converged to a solution of
$V_{gsr} = -41.67 - (0.84 \times b) - (0.014 \times
b^2)$.  This fit is shown as the green line in Figure~\ref{vvbsf2},
with the limits we chose for CPS velocity selection at $\pm20$ km
s$^{-1}$ on either side of this fit.  This is the velocity cut that
was used to select the red and blue points in
Figure~\ref{csf2}, which constitute our ``best'' sample of CPS stars.

\begin{figure}[!t]
\includegraphics[width=5.5in]{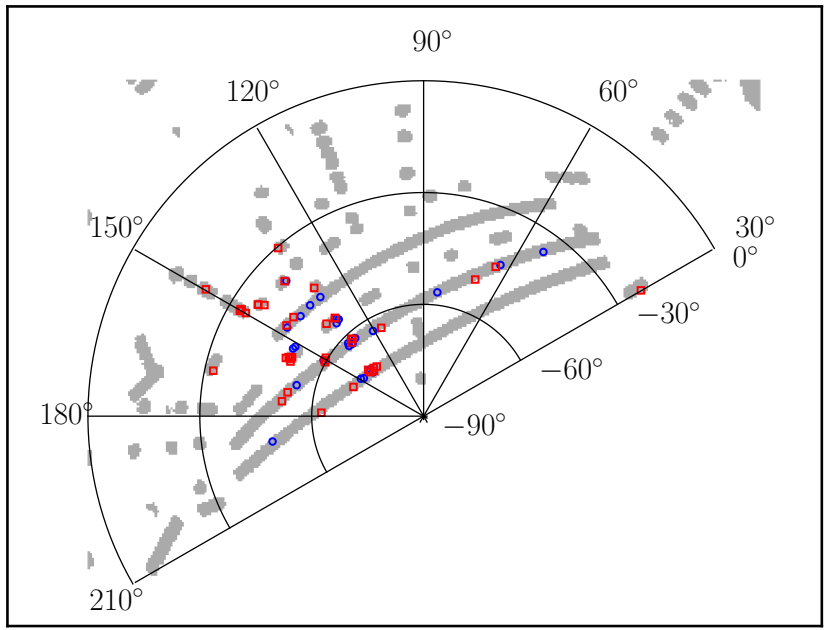}
\caption{Polar plot in Galactic coordinates, centered on the south Galactic
cap, of metal-poor ($-2.5 < $[Fe/H] $< -2.0$) CPS candidates selected
from SDSS DR8 to lie within the BHB and RGB color-magnitude
selections, as well as to have $V_{gsr}$ within 20 km s$^{-1}$ of the
trendline fit in Figure~\ref{vvbsf2}. Symbol colors and shapes are the same as in Figure~\ref{vvbsf2}. Nearly all of the stars thus
selected are in the region centered on $l \sim 145\arcdeg$ where CPS
is known to be located. The shaded area shows the footprint of SDSS DR8, from which the spectroscopic targets were selected.}
\label{lps2}
\end{figure}

We now check to make sure the spectroscopically selected stars match
our expectations for position on the sky.  Using this newly found
relationship, we select all of the stars from within the BHB and RGB color-$g_{corr}$ selection boxes, with velocities within 20
km s$^{-1}$ of the trendline, and that are also metal-poor ($-2.5 <$
[Fe/H] $< -2.0$), and plot them in a polar plot of the south Galactic
cap (Figure~\ref{lps2}). Here, we can see that the stars match
spatially with the photometric selection in Figure~\ref{lbppb}, with
very few stars matching the selection criteria outside of the CPS
region.

Because the stars with measured velocities are confined to ``clumps''
at the positions where SEGUE plates were located, a selection in the
same bins used to derive positions and distances would skew the
velocity measurement in each bin toward positions of highest
concentrations of data.  (See, for example, Figure~\ref{vvbsf2}. The
clump of stars at $b \sim -71\arcdeg$ would not fall within any of the
bins from Table~\ref{cpsk}, and the $-70\arcdeg < b < -62\arcdeg$ bin
would be skewed heavily toward the $b = -62\arcdeg$ end rather than
reflecting the velocity at $b = -66\arcdeg$.) Thus we chose to create
separate bins from the data in Figure~\ref{vvbsf2} with which to
measure the mean velocities, centered on the highest concentrations of
CPS candidates.  
Mean velocities and intrinsic velocity dispersions (accounting for the measurement errors) in each bin were calculated using a maximum likelihood method (e.g., \citealt{pm93,hgi+94,kwe+02}).
The mean latitude, $V_{gsr}$, intrinsic velocity dispersion,
and number of stars in each of these bins are given in
Table~\ref{mv}. We note that these velocity dispersions of $\sim 4-8$~km~s$^{-1}$ are typical of dwarf galaxies in the Local Group (see \citealt{2012AJ....144....4M} and references therein), and higher than typical dispersions in globular clusters.
To derive the velocities in Table~\ref{cpsk}, we
interpolate between the values in Table~\ref{mv}; for example, the
velocity at $b = -66\arcdeg$ in Table~\ref{cpsk} and its error were
linearly interpolated from the $b \sim -71\arcdeg$ and $b \sim
-61\arcdeg$ values in Table~\ref{mv}. Note also that all of the values
given in each of these tables are consistent with the polynomial fit
of $V_{gsr}$ vs. $b$ seen in Figure~\ref{vvbsf2}.

Finally, we note that we often find halo substructure in plots of line-of-sight velocity for
restricted volumes of space. There is a previously
unidentified clump of stars at $b=-38\arcdeg$ and $V_{gsr}=95$~km~s$^{-1}$ in Figures~\ref{vvbsm} and \ref{vvbsf2}.  We investigated this
clump, and discovered eight low surface gravity, low proper motion
stars that are within a degree of $(l,b)=(136\arcdeg,-38\arcdeg$),
have metallicities in the range $-2.5<$ [Fe/H] $<-1.5$, and follow a
giant branch very similar to the CPS giant branch.  This structure (also seen photometrically by \citealt{g12} and \citealt{bgk12}) is the subject of a
separate publication \citep{mcn+13}, in which we show that these stars are
part of a narrow tidal stream that we dub the ``Pisces Stellar Stream.''

\section{Fitting an orbit to the Cetus Polar Stream}\label{sec:orbit}

\begin{figure}[!t]
\includegraphics[height=4.5in]{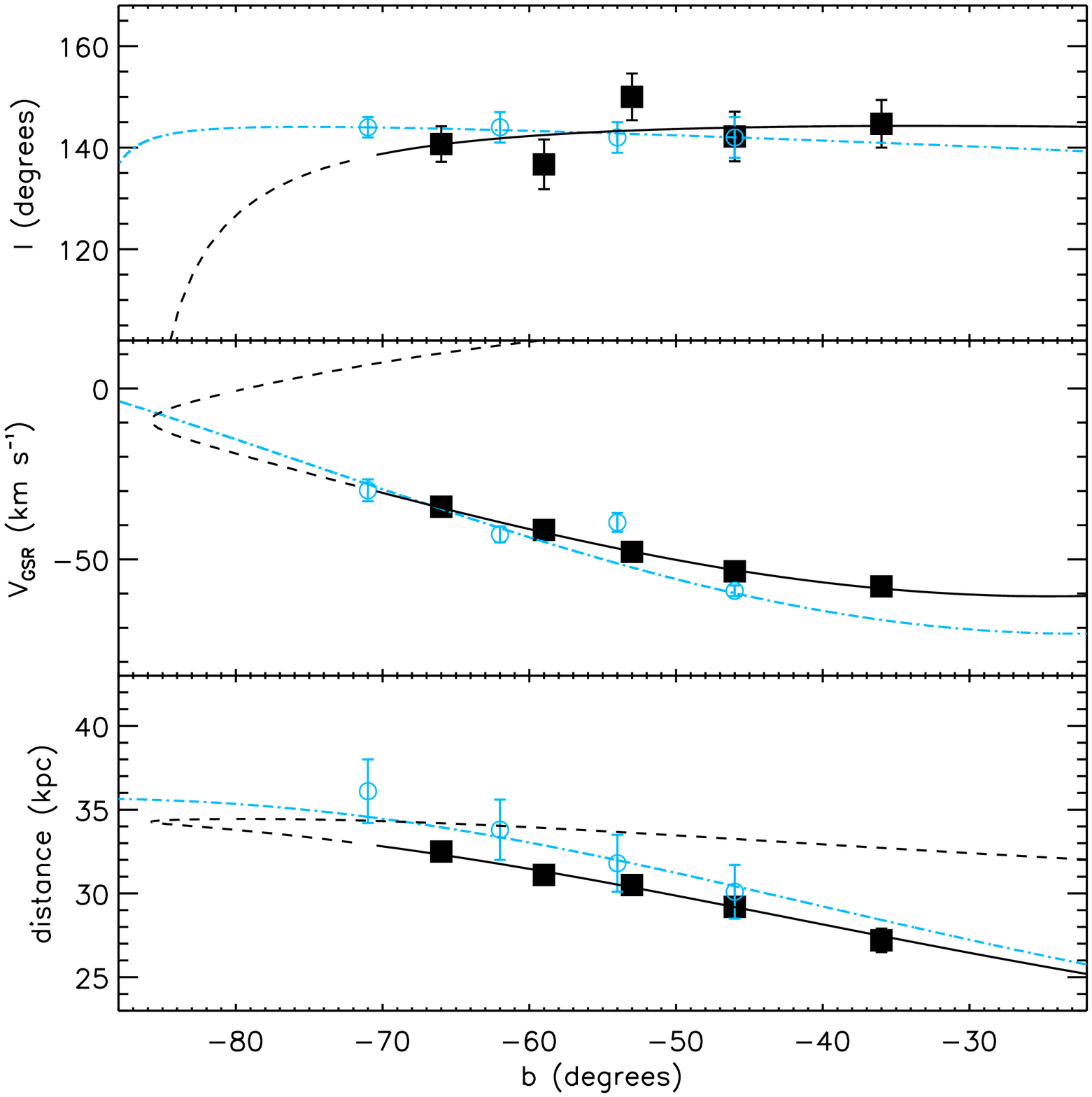}
\caption{Position, $V_{gsr}$, and heliocentric distance for the orbit fit to the five data points from Table~\ref{cpsk}, which are
shown as filled squares with their associated error bars (note that
the error bars in the middle and bottom panels are smaller than the
point size). Because the stream is at roughly constant Galactic
longitude, the data are shown as a function of Galactic latitude,
which corresponds roughly to position along the stream. The best-fit
orbit is given by the line in each panel, with the forward integration
from our selected fiducial point at $(l, b) = (138.2\arcdeg,
-71.0\arcdeg)$ shown as a solid line, and the backward integration as
a dashed line. The open circles (and their associated error bars)
represent the four data points used by \citet{nyw09} to fit the CPS
orbit; the point at $b \sim -46\arcdeg$ is obscured by the
solid square. The blue (dot-dashed) line is the best fit orbit from
\citet{nyw09}.}
\label{lbdv_orbit}
\end{figure}

Having determined the position, velocity, and distance trends along
the stream, we wish to use these to constrain the orbit of the Cetus
Polar Stream progenitor.  The orbit fitting routine requires a set of
discrete data points, rather than the general trends we have fit as a function of latitude for the CPS. 
Note that we fit an orbit assuming that the debris we have measured follows the orbit of the progenitor. This assumption that the tidal debris trace the progenitor's orbit is not strictly true, and the stream-orbit misalignment is essentially independent of progenitor mass \citep{sandersbinney13a}. 
We test and discuss this assumption that the stream follows the orbit in Section~\ref{sec:nbody}, which describes $N$-body simulations of this tidal stream.

\begin{figure}[!t]
\includegraphics[height=4.5in]{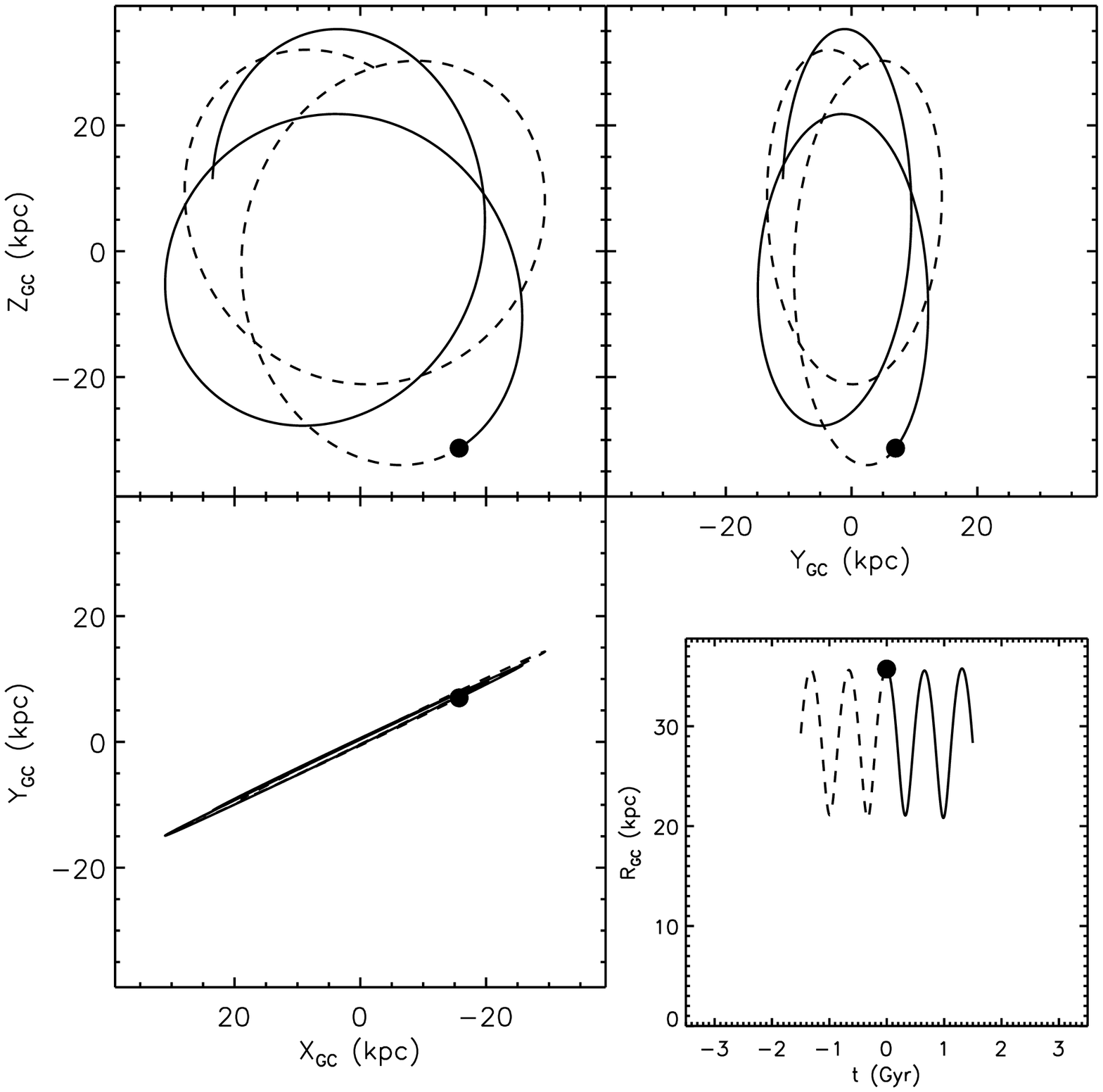}
\caption{Path of the best-fitting Cetus Polar Stream orbit in Galactocentric
$XYZ_{GC}$ coordinates (in a right-handed system, with the Sun at $(X,Y,Z)_{GC} = (-8, 0, 0)$ kpc). The large dot represents our fiducial point,
with forward integration shown as a solid line, and the backward
integration as a dashed curve. The nearly polar path of the
low-eccentricity ($e \sim 0.2$) orbit is evident in the lower left
panel, which shows the $XY_{GC}$ projection. The orbit is confined to
a narrow range in this plane. The lower right panel depicts the
Galactocentric distance of the orbit as a function of time. Evidently
the selected fiducial point is very near the apocenter of the orbit,
which occurred at $R_{GC} \approx 36$ kpc only $\sim2$ Myr prior to the
anchor point of our fitting.}
\label{xyz_orbit}
\end{figure}

\begin{deluxetable}{lrrrrrrrrrr}
\tabletypesize{\scriptsize}
\tablecaption{Cetus Polar Stream Kinematics and Spatial Positions \label{cpsk}}
\tablehead{\colhead{$b$ (deg)\tablenotemark{a}} & \colhead{$l$ (deg)} & \colhead{$\sigma_{\rm l}$ (deg)} & \colhead{$\sigma$ (kpc)}   & \colhead{N} & \colhead{$D_{\rm mod, fit}$} & \colhead{$\sigma_{D_{\rm mod, fit}}$} & \colhead{d$_{\rm fit}$ (kpc)} & \colhead{$\sigma_{\rm d, fit}$ (kpc)} &  \colhead{$V_{gsr}$ (km s$^{-1}$)}} 

\startdata
\\
-70:-62 & $140.7\pm3.5$ & $11.6\pm3.3$ & $2.7\pm0.8$ & 43 & $17.56\pm0.03$ & $0.09\pm0.03$ & $32.5\pm0.5$ & 1.3 & $-35.2\pm2.9$ \\
-62:-56 & $136.7\pm4.9$ & $5.5\pm3.7$ & $1.5\pm1.0$ & 37 & $17.46\pm0.03$ & $0.09\pm0.02$ & $31.1\pm0.5$ & 1.3 & $-42.2\pm2.6$\\
-56:-50 & $150.0\pm4.6$ & $15.1\pm5.1$ & $4.8\pm1.6$ & 44 & $17.42\pm0.03$ & $0.11\pm0.03$ & $30.5\pm0.5$ & 1.5 & $-47.8\pm2.3$\\
-50:-42 & $142.2\pm4.9$ & $5.0\pm2.6$ & $1.8\pm0.9$ & 35 & $17.33\pm0.03$ & $0.07\pm0.02$ & $29.2\pm0.4$ & 0.9 & $-53.0\pm2.2$\\
-42:-30 & $144.7\pm4.7$ & $9.8\pm4.2$ & $3.8\pm1.6$ & 40 & $17.17\pm0.06$ & $0.14\pm0.05$ & $27.2\pm0.7$ & 1.7 & $-57.9\pm2.4$\\
\\
\enddata
\tablenotetext{a}{This column represents a latitude range between which stars were selected.}
\end{deluxetable}

\begin{deluxetable}{lrrrr}
\tabletypesize{\small}
\tablecaption{Measured Velocities and Velocity Dispersions\label{mv}}
\tablehead{\colhead{$b$ (deg)} & \colhead{$V_{gsr}$ (km s$^{-1}$)} & \colhead{$\sigma_{V, 0}$ (km s$^{-1}$)} & \colhead{N}} 

\startdata
\\
-71.1 & $-29.8\pm3.0$ & $7.8\pm2.5$ & 10 \\  
-61.4 & $-40.0\pm2.7$ & $7.4\pm2.3$ & 12 \\ 
-48.8 & $-51.6\pm2.1$ & $6.7\pm2.1$ & 15 \\ 
-35.8 & $-58.0\pm2.4$ & $4.8\pm2.0$ & 9 \\  
\\
\enddata
\end{deluxetable}

The orbit was fit using the data at five positions given in
Table~\ref{cpsk} following the techniques described by
\citet{wnz+09}. The fit assumed a fixed Galactic gravitational
potential of the same form used by \citet{wnz+09}, which was in turn
modeled after the potential of \citet{ljm05} and \citet {jms+99}. This
model contains a three-component potential made up of disk, bulge,
and halo components. All parameters in the model were fixed at the same
values given in Table~3 of \citet {wnz+09}, and the Sun was taken to
be 8 kpc from the Galactic center. The technique was slightly improved
so that the orbit is not constrained to pass through the sky position
of any of the data points.

Orbits are uniquely defined by a gravitational potential, a point on
the orbit, and a velocity at that point. Because it doesn't matter
where along the orbit the velocity is specified, we are free to
arbitrarily choose one of the three spatial parameters.  For fitting,
we fixed the Galactic latitude of the point on the orbit at $b =
-71\arcdeg$, and fit five orbital parameters: the heliocentric
distance ($R$), Galactic longitude ($l$), and the components
$V_X, V_Y$, and $V_Z$ of Galactic Cartesian space velocity.  We evolve
the test particle orbit both forward and backward from the starting
position, and perform a goodness-of-fit calculation comparing the
derived orbit to the data points in Table~\ref{cpsk}. The best-fit
parameters are optimized using a gradient descent method to search
parameter space. We find best-fit parameters at $b = -71\arcdeg$ of $l
= 138.2\pm3.8\arcdeg$, $R = 32.9\pm0.3$ kpc, and $(V_X, V_Y, V_Z) =
(-118.1, 64.8, 76.3)\pm(7.2,10.3,3.0)$ km s$^{-1}$.

The errors in the measured orbit parameters\footnote{Note that the errors quoted on all orbital parameters here are the formal errors from the fitting. We show in Section~\ref{sec:nbody} that the assumption that the debris follows the progenitor's orbit is clearly not valid. The errors on the orbit parameters are thus somewhat arbitrary, since the true orbit is likely not the one arrived at by fitting with our naive assumptions. Nonetheless, we believe the values quoted here are a good first approximation to the true CPS orbit.} are smaller if the minimum
in the $\chi^2$ surface is narrower. These errors were calculated from
the square root of the diagonal elements of twice the inverse Hessian
matrix. The Hessian matrix consists of the second derivatives of
$\chi^2$ (not the reduced $\chi^2$) with respect to the measured
parameters, evaluated at the minimum.

The position, velocity, and heliocentric distance as a function of Galactic
latitude predicted by the integrated best-fitting orbit is shown in
Figure~\ref{lbdv_orbit}, with the data points constraining the fit
shown as large filled squares. This fit had a formal reduced $\chi^2$
of 0.48. This $\chi^2$ is rather low, and likely arises because we have conservatively estimated our uncertainties.
Note also that we have not discussed the
effects of possible systematic offsets in the distance scale. We tried
an additional orbit fit that included a multiplicative factor to scale
the distances as a free parameter; this fit did not improve the
$\chi^2$, so we retained the fit without the scale factor.

We find that the Cetus Polar Stream is on a rather polar orbit inclined to the
Galactic plane by $i \sim 87.0^{+2.0}_{-1.3}$ degrees. This orbit has an
apogalactic distance from the Galactic center of $35.67\pm0.01$ kpc and is at
$23.9^{+3.4}_{-3.0}$ kpc when passing through perigalacticon. This results in
an orbital eccentricity of $e = 0.20\pm0.07$.  The period of our
derived CPS orbit is $\sim0.694\pm0.04$ Gyr ($\sim694$ Myr), measured
between consecutive pericentric passages. Errors on all of these quantities were derived by integrating the orbits using the maximum and minimum possible total velocities from the orbit-fitting errors in ($V_X, V_Y, V_Z$) and comparing these to the best fit orbit. The orbit is shown in
Galactic Cartesian $XYZ_{GC}$ coordinates\footnote{The Galactic $(X, Y, Z)_{\rm GC}$ coordinates refer to a right-handed Cartesian frame with origin at the Galactic center, $X_{\rm GC}$ positive in the direction from the Sun to the Galactic center, $Y_{\rm GC}$ in the direction of the Sun's rotation through the Galaxy, and $Z_{\rm GC}$ out of the plane toward the north Galactic hemisphere (i.e., $b>0\arcdeg$). Assuming that the Sun is
8.0~kpc from the Galactic center, it has coordinates of $(X, Y, Z)_{\rm GC} = (-8.0,0,0)$~kpc.} in Figure~\ref{xyz_orbit},
integrated for a total of 1.5 Gyr (0.75 Gyr each in the ``forward''
and ``backward'' directions from our chosen position). The forward
integration is given by the solid lines, and the backward integration
is the dashed lines, as in Figure~\ref{lbdv_orbit}.  The bottom right
panel shows the Galactocentric distance as a function of time; from
this panel it is clear that our chosen position to anchor the orbit is
very near the apocenter of the orbit. Indeed, we find that the nearest
apogalacticon was only $\sim2$ Myr prior to this fiducial point, at
a position of $(l, b) = (137.8\arcdeg, -71.6\arcdeg)$. The most recent pericentric passage was
$\sim350$ Myr ago, at a position of  $(l, b) = (337.4\arcdeg, 7.0\arcdeg)$, and a distance of $\sim27.2$ kpc from the Galactic center.
Note that our assumption that the debris trace the orbit likely leads to differences in the derived orbital parameters at the $\sim10\%$ level, especially for a massive progenitor. This will be discussed further in the next section.

%%%%%%%%%%%%%%%%%%%%%%%%%%%%%%%%%%%%%%
\section{N-body model} \label{sec:nbody}

We now use the orbit we have derived in combination with information about the velocities, velocity dispersions, stellar density, and three-dimensional positions of observed CPS 
debris to explore the nature of the stream's (unknown) progenitor. To 
do so, we use $N$-body simulations of satellites disrupting in a Milky Way-like gravitational 
potential, on the well-defined orbit we have measured in this work. The density distribution of 
debris along the stream, as well as the velocity dispersion, stream width, and line-of-sight depth, are sensitive to the 
mass and size of the progenitor, as well as the time that the progenitor has been orbiting the Milky Way. A comprehensive modeling effort is beyond the scope of this work; here 
we aim to gain insight into the nature of the progenitor that will inform future, more detailed, 
modeling efforts.

The $N$-body simulations were run using the gyrfalcON tool (Dehnen 2002) of the NEMO Stellar Dynamics Toolbox (Teuben 1995). Satellites with masses configured using a Plummer model (Plummer 1911) were evolved in the same Galactic gravitational potential used to derive orbits. Each simulation has $10^4$ bodies.  

We started out with the assumption that mass follows light, so that the distribution of masses from the $N$-body simulation has the same density and 
velocity profile as the observed stars.  We note that this assumption has been shown to work in modeling dwarf spheroidals by Mu{\~n}oz et al. (2008). 
With this assuption, the Plummer scale radius of each satellite was chosen, roughly following the scaling relations from Tollerud et al. (2011; see, e.g., 
their Figure~7). A Plummer radius of 1~kpc was used for the $10^8 M_{\Sun}$ satellite, with the remaining radii scaled using the empirical relation 
(Tollerud et al. 2011) between the half-light radius and the dark matter mass within that radius: $M_{1/2}^{\rm DM} \propto r_{1/2}^{2.32}$ (note that we 
assumed that the Plummer radius is roughly equal to the half-light radius).  Since we assume mass follows light, the radius of the dark matter and the 
radius of the luminous matter are the same. 

The total masses of the Plummer spheres were varied from $10^5$ to $10^9$ solar masses, with Plummer scale radii of $a_p \sim (50, 150, 400, 1000, 
2500)$ pc corresponding to $(10^5, 10^6, 10^7, 10^8, 10^9)~M_{\Sun}$ models, respectively (note that our choice of scaling differs little from a simple 
choice of $r \propto M^{1/3}$, which would produce satellites with the same mass density).

\begin{figure}[!t]
\includegraphics[width=4.5in]{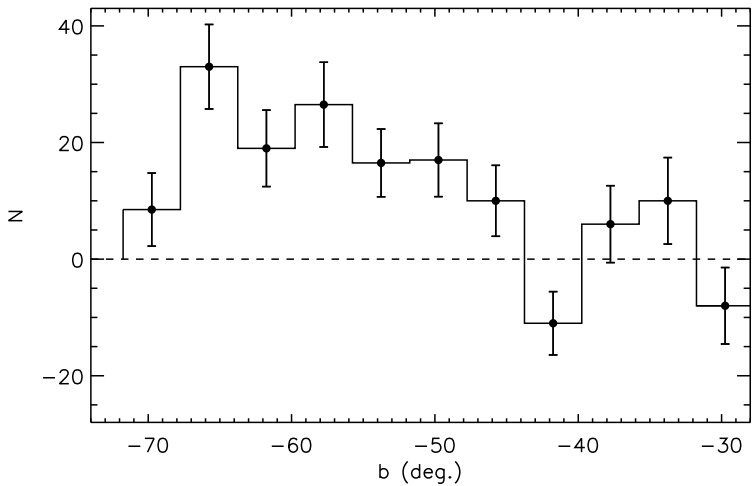}
\caption{Density of candidate photometrically-selected Cetus Polar Stream BHB stars selected using the photometric cuts discussed earlier, and limited to a distance modulus range of $16.5 < D_{\rm mod} < 18.0$ to bracket the CPS debris distances. Each bin represents the difference in number counts of BHB stars between an ``on-stream'' field between $130\arcdeg < l < 152\arcdeg$ and an ``off-stream,'' background field selected from $86\arcdeg < l < 130\arcdeg$ (scaled down by the factor of two difference in area). Error bars are derived from the Poisson errors in the number counts from each sample. A clear trend of decreasing BHB density along the stream can be seen between $-70\arcdeg < b \lesssim -40\arcdeg$. Above $b \sim -40\arcdeg$, the residuals are consistent with zero (i.e., the stream does not contain excess BHB stars above the background level), though from 
Figure 5 we expect there are at least a few stream stars with $b>-42\arcdeg$.
}
\label{fig:bhbcounts}
\end{figure}

We have seen that the density of stars in the CPS falls as one approaches the Galactic plane. This is also along the direction of our measured orbital motion, so 
that the observed stars near the south Galactic cap must be the leading tidal tail of a satellite (which could be completely disrupted) that is not too far 
behind the observed stars on the orbit.  We place a point mass at the position $(l,b) = (138.2\arcdeg, -71\arcdeg)$ on the orbit at a distance of 32.9 kpc, 
then integrate the point backwards on the orbit for 3 Gyr.  At that new position, we place a Plummer sphere of bodies representing the dwarf galaxy with 
the bulk velocity of the forward orbit at that position. We then integrate the $N$-body forward for three Gyr, just over four full orbits, so the dwarf galaxy ends 
up approximately at our initial position.

\begin{figure}[!t]
\includegraphics[height=4.0in]{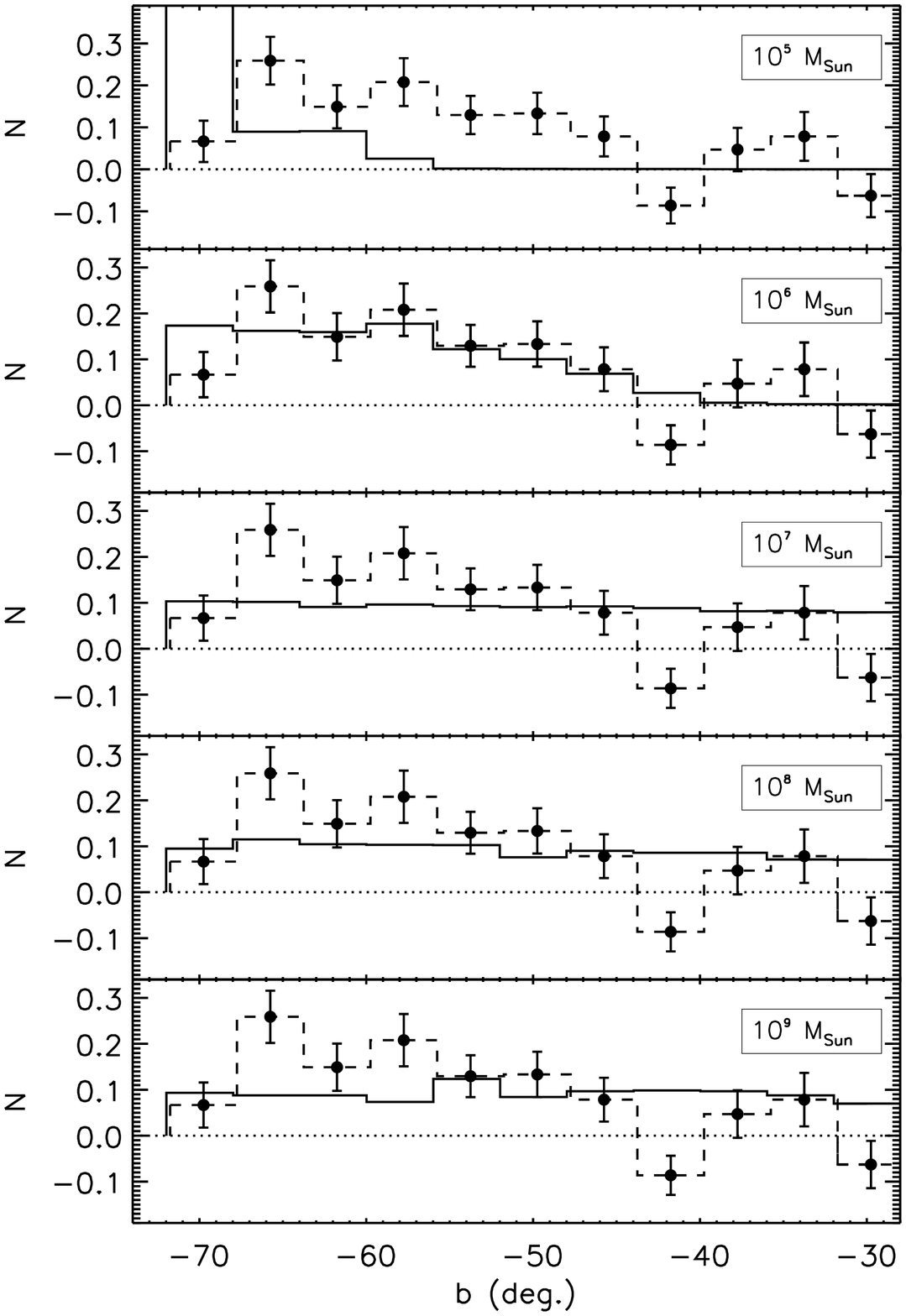}
\includegraphics[height=4.0in]{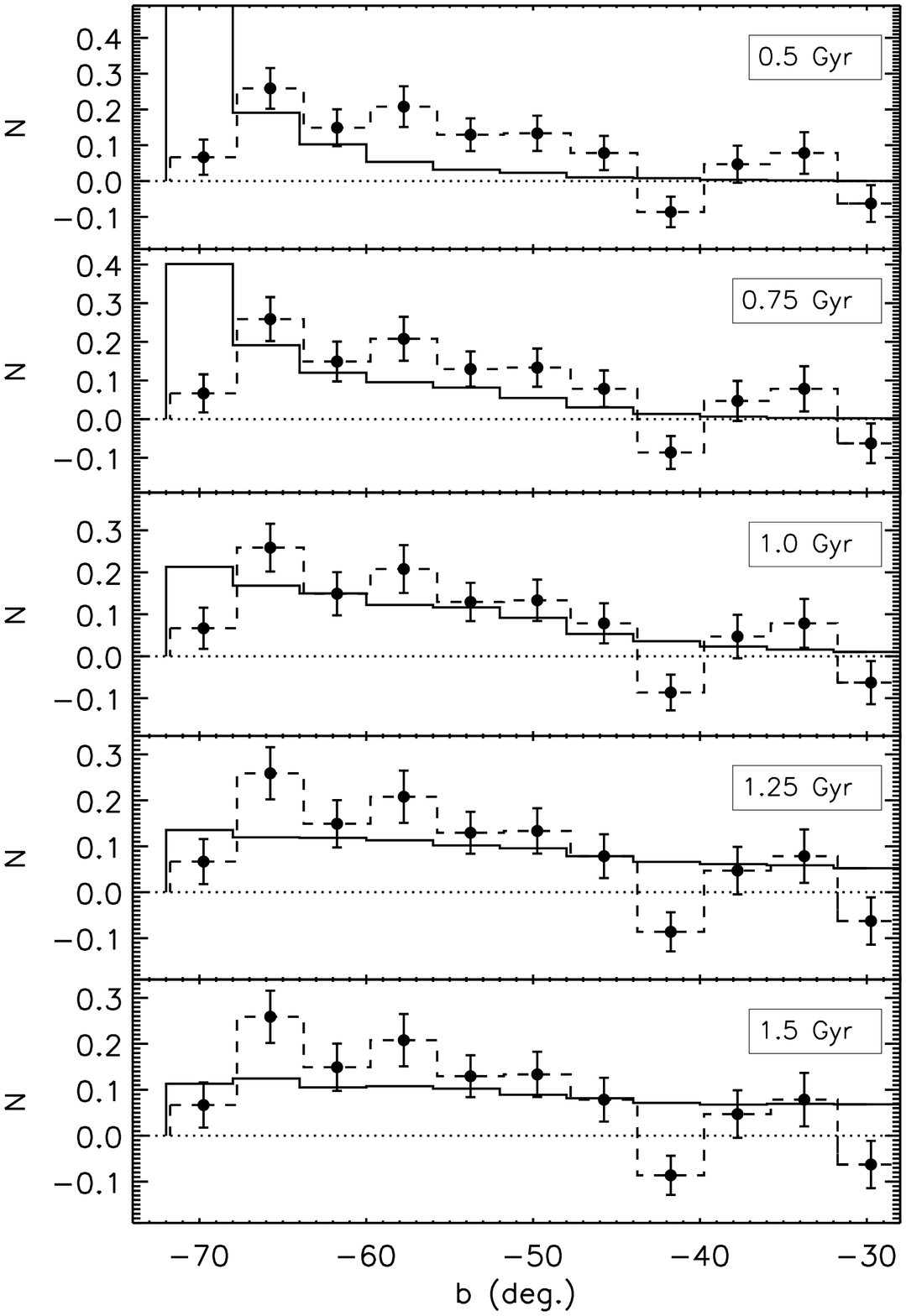}
\caption{Normalized density of tidal debris from each of the $N$-body models of satellites on the best-fitting Cetus Polar Stream orbit. The debris are selected and binned in the same way as the BHB stars in Figure~\ref{fig:bhbcounts} to facilitate direct comparison. The normalized density of BHB stars from Figure~\ref{fig:bhbcounts} is overplotted in each panel as a dashed line with associated error bars. Each bin (for both the observational data and the model) has been normalized by the total number of points in the sample. From top to bottom, the left panel shows models with progenitor masses of $10^5, 10^6, 10^7, 10^8$, and $10^9 M_{\Sun}$ evolved for 3 Gyr. The progenitor (or its remains) is just off the plot near the south Galactic pole. The right panels show the $10^8 M_{\Sun}$, 1 kpc radius model evolved for 0.5, 0.75, 1.0, 1.25, and 1.5 Gyr (top to bottom). In the 3 Gyr models, the least massive ($10^5 M_{\Sun}$) satellite has suffered little tidal disruption. Its debris is not spread far along the orbit because the satellite is compact and has very low velocity dispersion; this stream is thus more sharply peaked than the observed density. The $10^7-10^9 M_{\Sun}$, 3 Gyr satellites show an almost constant density of debris along the entire length of the region plotted, while in the $10^6 M_{\Sun}$ progenitor, the fall-off in stream density is more rapid as a function of latitude (essentially as a function of angular distance along the stream, since the stream runs along nearly constant longitude). The $10^6 M_{\Sun}$ model is most consistent with the stellar density of the CPS along latitude, while the additional information on stream stars suggests a more massive progenitor. In the right panels, the $10^8 M_{\Sun}$, 1 Gyr model matches the data best. This would require that a fairly massive satellite had its (previously different) orbit perturbed 1 Gyr ago to place it on the nearly circular CPS orbit for only $\sim1.5$ orbits.
}
\label{fig:nbodycounts}
\end{figure}

We measure the width across the stream ($\sigma_l$, which is the $\sigma$ of a Gaussian fit to the profile), the line of sight depth of the stream ($
\sigma_d$; again from a Gaussian fit), and the stream velocity dispersion ($\sigma_{\rm V}$), from the results of each $N$-body model. Table~\ref{tab:cpsnbody} shows the measured values for the same five latitude bins we used to analyze the data, which should be compared with the 
observed properties of the stream stars in the same table.  The stream widths, $\sigma_l$ and $\sigma_d$, both come from photometrically selected 
BHB stars, while the velocity dispersion includes BHBs and RGB stars. The width in Galactic longitude, $\sigma_l$, is large for the BHB stars, ranging 
from $5.0\arcdeg$ to $15.1\arcdeg$, and has large uncertainties. This likely arises because of the low numbers of BHB stars in the samples, such that the 
poorly-defined background BHB density broadens the wings in some of these fits. 

The $10^5 M_{\Sun}$, $a_p = 50$ pc progenitor, which has short, narrow debris tails resembling those of the Pal 5 globular cluster (e.g., Grillmair \& 
Dionatos 2006; Odenkirchen et al. 2001, 2003; Rockosi et al. 2002), gives unreliable results for many of our measurements simply because there is 
little debris to be measured.  Nevertheless, it is clear that the low-mass ($M_{\Sun} < 10^7$) progenitors produce streams much narrower in both width 
on the sky and line-of-sight depth, than observed in the CPS. The low mass progenitors also have velocity dispersions that are too small.  The measured 
width and depth of the stream stars suggest a $10^8-10^9 M_{\Sun}$ progenitor mass, while the velocity dispersions suggest a $10^7-10^8 M_{\Sun}$ 
progenitor mass.
Thus it seems that the physical widths (both on the sky and along the line of sight) and the velocity dispersions we have measured are telling us that the 
progenitor of the Cetus Polar Stream had a mass of $\sim10^8 M_{\Sun}$.

However, we now show that the density of stars along the stream is inconsistent with the mass-follows-light progenitor in our model.
We use the BHB stars selected photometrically by the methods described earlier in this 
work. The BHB sample is restricted to distance moduli of $16.5 < D_{\rm mod} < 18.0$ to choose only the distance range containing CPS debris. The 
entire available range of Galactic latitude is chosen ($-72\arcdeg < b < -28\arcdeg$). We select an ``on-stream'' and ``off-stream,'' or background, 
sample. The on-stream BHB stars are selected in the longitude range of $130\arcdeg < l < 152\arcdeg$ to include only the regions completely sampled 
by SDSS DR8. The off-stream sample spans $86\arcdeg < l < 130\arcdeg$ over the same latitudes. We bin the number counts in 4-degree bins of 
latitude for each sample, then scale down the number counts in the off-stream regions by a factor of two to account for the additional area sampled. 

Figure~\ref{fig:bhbcounts} shows the density profile of BHB stars along the CPS after subtraction of the ``off-stream'' background density. There is clearly an excess population of BHB stars in the CPS 
compared to the background at latitudes $-70\arcdeg < b \lesssim -40\arcdeg$. Above $b \sim -40\arcdeg$, the excess seems to disappear, with the 
BHB numbers consistent with the adjacent background level.  However, recall from Figure 5 that there must be at least a few stream stars with $-42\arcdeg < b < -30\arcdeg$.  This density gradient starting from highest density near the south Galactic cap and 
decreasing toward the Galactic plane must be reproduced by any mass-follows-light model of the stream.  Note that the first bin in the figure is low because the area of sky between two longitudes near the south Galactic pole is small, and the stream does not go exactly through the pole; the density is actually highest in the bin nearest the pole, since that is close to the final position of the satellite.  

We measure the density profile as a function of latitude for the set of five $N$-body models that have the progenitor ending up near the south Galactic 
cap. The debris was selected from the same longitude and distance ranges used for the BHB density in Figure~\ref{fig:bhbcounts} and histogrammed 
in 4-degree bins as in the BHB figure. These density profiles are displayed in the left panels of Figure~\ref{fig:nbodycounts} as solid lines for each of the five progenitor 
masses. Number counts in each bin were normalized by the total number of points in each panel to facilitate comparison with the BHB density profile, 
which is overlaid as a dashed histogram (with error bars). The $10^5 M_{\Sun}$ progenitor has suffered little tidal disruption, and is thus strongly 
peaked near the satellite; this is likely because of the small (50 pc) radius of the satellite. The largest satellites (0.4-2.5 kpc, $10^7-10^9 M_{\Sun}$) produce nearly constant density along latitude, indicating that they are strongly disrupted and that debris has spread extensively along the orbit.  Note that although our data is near the apogalacticon of the orbit, the orbit is not very eccentric so highly disrupted satellites have only a small increase in stellar density at the position in the stream that is most distant from the Galactic center.  The stellar density along the stream in the simulations with a more massive dwarf galaxy does not 
agree with the observed gradient in the measured BHB density. 

On the other hand, the $10^6 M_{\Sun}$ model density appears to agree very well with the observations, with all five of the bins between $-64\arcdeg < b < 
-44\arcdeg$ in agreemeent with the BHB number counts within 1$\sigma$. However, we have already shown that a satellite with $10^6 M_{\Sun}$ is not consistent with the width, depth and velocity dispersion of the stream.

\begin{deluxetable}{rrrrrrrrrr}
\tabletypesize{\scriptsize}
\tablecaption{Cetus Polar Stream N-body results (with measured values for comparison) \label{tab:cpsnbody}}

\tablehead{
& & \multicolumn{3}{c}{3-Gyr models, varying mass}  & &
\multicolumn{4}{c}{0.5-1.5 Gyr models, $M=10^8 M_{\Sun}$}  \\
\cline{1-5}   \cline{7-10}

\colhead{mass} &
\colhead{$b$\tablenotemark{a}} & 
\colhead{$\sigma_l$} & 
\colhead{$\sigma_{\rm d}$} & 
\colhead{$\sigma_{\rm V}$} & &
\colhead{time} & 
\colhead{$\sigma_l$} & 
\colhead{$\sigma_{\rm d}$} & 
\colhead{$\sigma_{\rm V}$} \\
\colhead{($M_{\sun}$)} &
\colhead{(deg)} & 
\colhead{(deg)} & 
\colhead{(kpc)} & 
\colhead{(km s$^{-1}$)} & &
\colhead{(Gyr)} &
\colhead{(deg)} & 
\colhead{(kpc)} & 
\colhead{(km s$^{-1}$)}
}

\startdata
\\
% mass  b   sig_l    sig_d    sig_v  
 observed & -70:-62 & $11.6\pm3.3$ &  $1.3^{+0.4}_{-0.5}$ & $7.6\pm2.4$ & & & & &\\
 - & -62:-56 & $5.5\pm3.7$ &  $1.3^{+0.2}_{-0.5}$ & $7.3\pm2.2$ & & & & & \\
 - & -56:-50 & $15.1\pm5.1$  & $1.5^{+0.4}_{-0.4}$ & $6.9\pm2.2$ & & & & &\\
 - & -50:-42 &  $5.0\pm2.6$  & $0.9^{+0.3}_{-0.3}$ & $6.3\pm2.1$ & & & & & \\
 - & -42:-30 & $9.8\pm4.2$  & $1.7^{+0.6}_{-0.6}$ & $4.9\pm2.0$ & & & & & \\ \\

\tableline \\

 $10^5$ & -70:-62 &   1.0  &   0.4 &  3.2 & & 0.50 &    4.8    &   2.4 &    14.8\\
 - & -62:-56 &  0.6 &  0.3 & 1.8 & & - & 4.8 &  1.5   &    7.3\\
 - & -56:-50 &  ... &  ... &  1.8 & & - &     4.7    &  1.5    &  6.3\\
 - & -50:-42 &  0.5  &  ... &  2.3 & & - &      4.3  &  1.1  &     6.9\\
 - & -42:-30 & ... &  ...  & 0.3 & & - &      4.1    &   0.2   &   14.5\\ \\

 $10^6$ & -70:-62 & 1.2 &   0.6 &  4.0 & &  0.75 &      5.1 &      1.9 &     11.3 \\
 - & -62:-56 &    1.3 &       0.4 &       2.9 & &  -        &      4.8  &    1.1   &    7.3\\
 - & -56:-50 &      1.2 &     0.4 &      2.7 & &  -         &      4.5  &     0.9  &     6.0\\
 - & -50:-42 &     0.8 &      0.6  &     2.6 & &  -         &      4.2  &     0.8  &     7.1\\
 - & -42:-30 &     0.9 &      0.5  &     10.0 & &  -      &       3.8  &     0.9  &    11.2\\ \\

 $10^7$ & -70:-62 &  2.6 &      0.6 &    4.8 & &  1.00 & 4.7 &      1.3 &   9.4\\
 - & -62:-56 &    2.3 &      0.5 &      4.4 & &  - &     4.3     &  0.7      &  9.0\\
 - & -56:-50 &    2.0 &      0.6 &      4.6 & &  - &       3.9     &  1.1    &  12.9\\
 - & -50:-42 &    1.9 &      0.7 &      4.6 & &  - &       3.9    &   1.8    &  13.8\\
 - & -42:-30 &    1.8 &      0.9 &      4.4 & &  - &       3.8     &  2.2    &  15.3\\ \\

 $10^8$ & -70:-62 &  5.6 &  0.9 & 9.5 & & 1.25 &     5.4  &      1.2   &    9.9\\
 - & -62:-56 &      5.4 & 0.9 & 9.2 & &   - &      4.2     &  1.1   &    9.7\\
 - & -56:-50 &     4.4 &  1.0 & 9.0 & &   - &       3.6    &   1.1  &     9.6\\
 - & -50:-42 &   4.1 & 1.0 & 9.3 & &      - &       3.6    &   1.2  &    10.2\\
 - & -42:-30 &     3.4 & 1.2 & 9.0 & &    - &       3.4    &   1.3  &     8.9\\ \\

 $10^9$ & -70:-62 & 14.4 & 2.1 & 20.6 & &  1.5 &      4.9  &     1.1  &     9.5\\
 - & -62:-56 &   17.0 & 2.3 & 20.9 & &  -     &  3.9     & 1.0    & 10.1\\
 - & -56:-50 &    9.9 & 2.0 & 23.4 & & -     &  4.0      & 1.1     & 10.2\\
 - & -50:-42 &    10.4 & 2.4 & 23.4 & &  -   &    3.4    &   1.1    &   8.8\\
 - & -42:-30 &    6.7 & 2.4 & 20.8 & &  -    &   3.1     &  1.1     &  7.4\\ \\

\\
\enddata
\tablenotetext{a}{This column represents a latitude range between which stars were selected.}
\tablecomments{For the 3-Gyr $N$-body models, particles were configured in Plummer spheres with scale radii of $(50, 150, 400, 1000, 2500)$ pc, respectively, for the models with masses of $(10^5, 10^6, 10^7, 10^8, 10^9)~M_{\Sun}$.}
\end{deluxetable}

\begin{figure}[!t]
\includegraphics[width=3.25in]{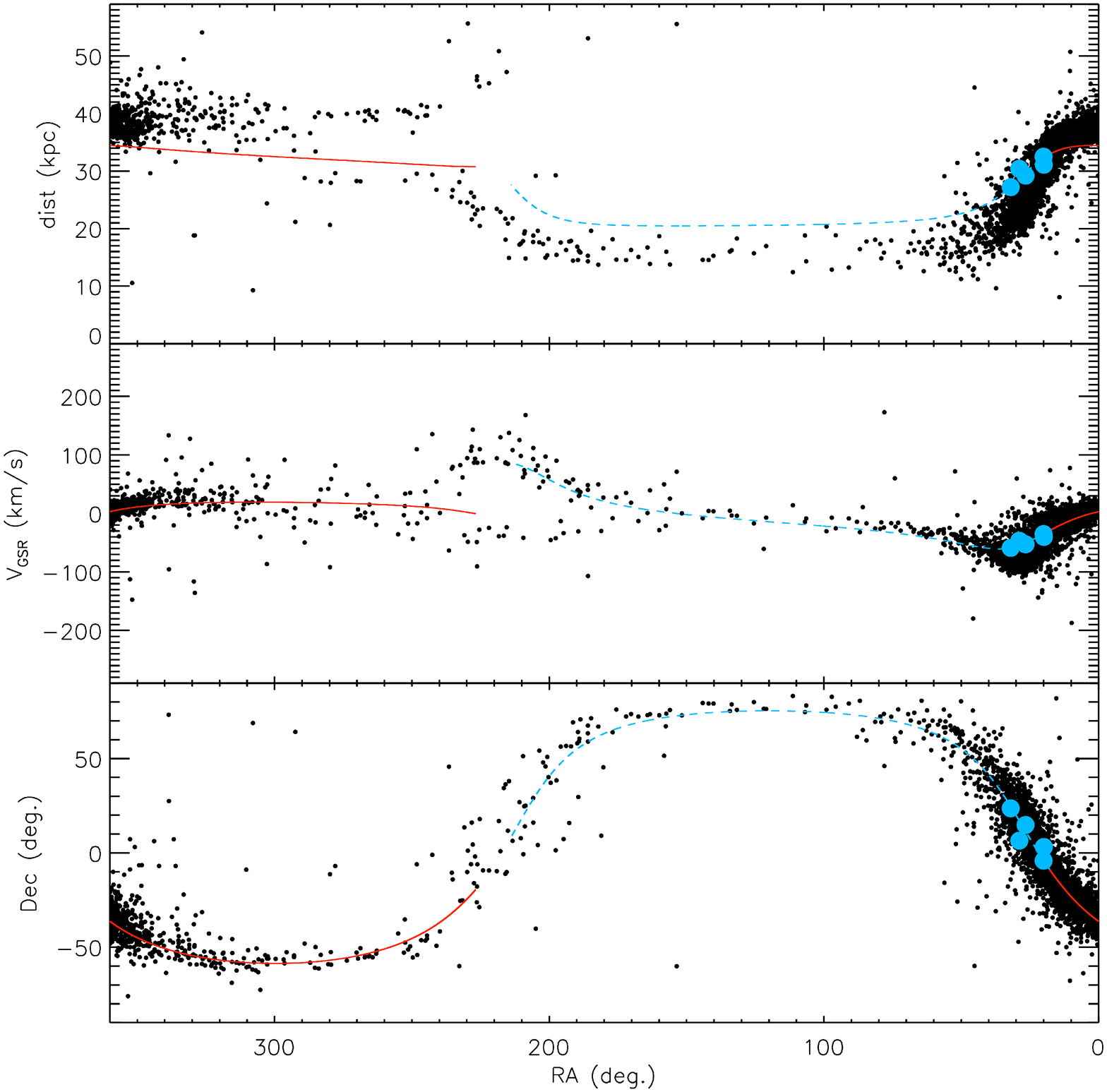}
\includegraphics[width=3.25in]{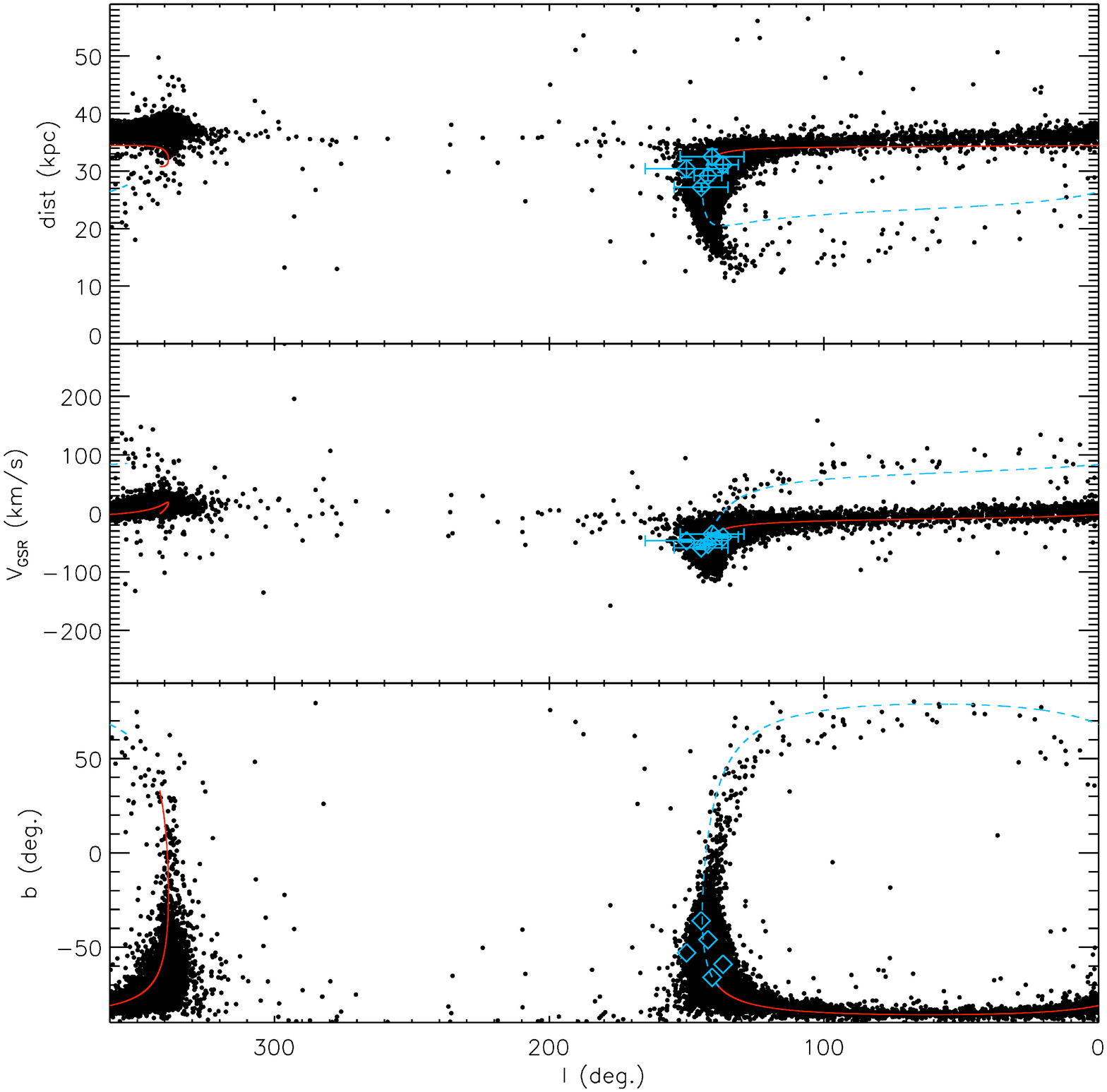}
\caption{Results of an $N$-body model of a satellite on the best-fitting Cetus 
Polar Stream orbit. The satellite was intialized as a Plummer sphere of 10,000 particles, with a total mass of $10^8 M_{\Sun}$. Its evolution was evolved for 1.0~Gyr in a model Milky Way potential. The black dots above are the model results. The panels show the distance from the Sun, $V_{\rm GSR}$, and position (right ascension and declination in the left column, and Galactic coordinates in the right column) of model particles. Overlaid solid red lines are the forward-integrated orbit, and dashed (blue) lines the reverse integration 
from our chosen fiducial position. Large blue points show the five data points derived in this work.
}
\label{fig:nbodyplots}
\end{figure}

Since the $10^8$ solar mass progenitor seemed a reasonable fit to the stream width and velocity dispersion, but the distribution of stars along the stream was too dispersed, we tried reducing the time the stream has been disrupting in an attempt to fit all of the measured quantities.  The right panels of Figure~\ref{fig:nbodycounts} show the distribution along the stream for disruption times of 0.5 to 1.5 Gyr, and Table~\ref{tab:cpsnbody} shows the velocity dispersion and stream widths for these simulations.  We find that a $10^8$ solar mass progenitor with a disruption time of 1-1.25 Gyr is a reasonable but not perfect fit to all of our data.  It is a bit difficult to fit the large apparent width of the stream with $N$-body simulations, even with a relatively large mass progenitor of $10^8 M_{\Sun}$. 
Note that the best-matching disruption time of $\sim1$ Gyr is about one and a half orbits of the satellite, and would require that the dwarf galaxy was perturbed into its present, nearly circular orbit fairly recently; we did not model the effects of a recent deflection of the satellite in the simulation.  It might be possible to adjust the disruption time slightly by varying the final position of the progenitor, but this would only produce a slight change to the best fit time.

The results of the 1 Gyr $N$-body simulation for a $10^8 M_{\Sun}$ satellite are shown in Figure~\ref{fig:nbodyplots}. This figure shows the distribution of model debris in distance, velocity, and position (right ascension and declination in the left column, and Galactic coordinates in the right panels).  One difficulty that is illuminated in Figure~\ref{fig:nbodyplots} is that our orbit, which was fit to the tidal debris, does not lead to an $N$-body simulation that
fits the measured stream distances for a massive 
($10^8 M_{\Sun}$) progenitor (in particular, see the upper left panel of Figure~\ref{fig:nbodyplots}).  The discrepancy between orbits fit to streams and the actual orbits of massive progenitors has been
highlighted by \citet{binney08} and \citet{sandersbinney13a}.

To address this discrepancy, we tried adjusting the orbit by hand to match the data.  We were able to plausibly match the $N$-body simulations by increasing the distance of the orbit from the Galactic center.  The new orbit has an apogalacticon of 38 kpc, a
perigalacticon of 26 kpc, an inclination of $88\arcdeg$, eccentricity of 0.19, and a period of 750 Myr.  All of these values are within 10\% of the orbit that was fit to the stream, and other than a shift to $\sim2$~kpc larger Galactocentric distances, all of the parameters are within the uncertainties of the orbital parameters from our original fit.  The evolution time of the simulation also affects the results; this particular orbit was a better fit to
the data with a 3 Gyr evolution time than a 1 Gyr evolution time.  
It is a much more difficult problem, and beyond the scope of this current effort, to fit orbital parameters by matching data directly to $N$-body simulations. Thus we cannot verify that our solution is unique or is in fact the best solution.  However, we note that this slightly altered orbit has very similar width, depth, velocity dispersion, and density along the stream as the original $N$-body simulation.  The best fit model is still a 
$10^8 M_{\Sun}$ satellite that may have been deflected into its current orbit $\sim 1$ Gyr ago.

Another more plausible solution is that mass does not follow light in the stream progenitor.  So far, our simulations have assumed that the $N$-body points, which are simply massive particles (and thus more representative of the {\it dark 
matter} in the satellite), will be distributed similarly to the luminous matter (i.e., the stars).  There is little reason to believe this is the case, and in fact the observed properties of many dwarf galaxies are often interpreted as showing evidence of significant dark matter halos in these objects (see, e.g., \citealt{2012AJ....144....4M} and references therein).  Our $N$-body comparisons seem to suggest that the CPS progenitor was a dwarf galaxy with a large mass-to-light ratio and a dark matter halo that extends beyond the distribution of stars.  A model of this type might explain the apparently small number of stars given the large spatial and velocity dispersion of the stream, and the relatively recent disruption of the stellar component of the progenitor as determined by the relatively steep gradient in star density along the stream.

Another argument in favor of an ultrafaint dwarf progenitor comes from the measured metallicity.  We showed in Sections~\ref{sec:feh} and \ref{sec:rgb} that the metallicity of CPS stars is sharply peaked between $-2.5 <$ [Fe/H] $< -2.0$. Assuming that we are looking at a disrupted dwarf galaxy, the $\langle[$Fe/H$]\rangle$ vs. M$_{\rm V}$ relation for Local Group dwarf galaxies of Fig.~12 of McConnachie 
(2012) suggests that a dwarf spheroidal with $\langle[$Fe/H$]\rangle \sim -2.25$ should have M$_{\rm V} \sim -6$. Likewise, the metallicity-luminosity relationship for Local Group dwarf galaxies from Simon \& Geha (2007) predicts ultra-faint dwarf spheroidals with $-4 >$~M$_{\rm V}~> -9$ should have mean metallicities between $-2.5 <$ [Fe/H] $< -2.0$, lower than most globular clusters at similar luminosities.  Thus the metallicity of the stars in the CPS is also consistent with its progenitor being an ultra-faint dwarf galaxy.

The velocity dispersion and stream width alone (without the $N$-body models for comparison) are too large for a globular cluster progenitor.  We therefore conclude that the progenitor was a dwarf galaxy.  One possible solution is a mass-follows-light progenitor of about $10^8$ solar masses, that was deflected into a nearly circular orbit of order one Gyr ago.  A more plausible solution is that the progenitor was a dark matter dominated, ultrafaint dwarf galaxy.  

Modeling a two component (dark matter plus stars) dwarf galaxy progenitor for the Cetus Polar Stream is beyond the scope of this work, but the prospect that the dark matter distribution of the progenitor dwarf galaxy could be encoded in the distribution of stars in a tidal tail is tantalizing, and will be pursued in a future paper.

We have been building the capacity to fit many parameters in an $N$-body simulation of dwarf galaxy tidal disruption using the Milkyway@home volunteer computing platform (\url{http://milkyway.cs.rpi.edu/milkyway/}).  We plan to eventually use this platform to find the best fit model parameters for the dwarf galaxy progenitor of the CPS given the available data for the tidal stream, and also to model other tidal streams in the Milky Way.

%%%%%%%%%%%%%%%%%%%%%%%%%%%%%%%%%%%%%%%%
\section{Conclusion}

In this work, we use BHB and RGB stars from SDSS DR8 to improve the determination of the Cetus Polar Stream orbit near the south Galactic cap. We then model the evolution of satellites on this orbit to assess the nature of the progenitor that produced the CPS. 

Much of the region of sky in which the CPS is located is dominated by the Sagittarius stream. However, we show that stars with the velocities of the Sgr tidal stream in the
region near the CPS are predominantly blue stragglers (BS), and that Sgr is virtually devoid of BHB stars.
This is in stark contrast with the CPS, which has a well populated BHB
and very few, if any, observed BS stars.
We are able to trace the CPS with BHB stars over more than 30 degrees in Galactic latitude along its nearly constant path in Galactic longitude. We were also able to disentangle CPS from Sgr based on the different velocities and mean metallicities of the two structures (see, e.g., Figure~\ref{vvbsm}). 

We fit an orbit to the Cetus Polar Stream tidal debris using the positions, distances, and velocities we have measured over at least $30-40\arcdeg$ of the stream's extent. Under the assumption that the orbit follows the tidal debris, the best fit CPS orbit has an eccentricity of $e = 0.20\pm0.07$, extending to an apogalactic distance from the Galactic center of $R_{\rm GC} = 35.7$~kpc at a point very near the lowest-latitude measurement in our study. At its perigalactic passage, the orbit is $\sim24$~kpc from the Galactic center. This orbit is inclined by $87\arcdeg$ to the Galactic plane, and has a period of $\sim700$~Myr.  Our $N$-body models indicate a high mass progenitor ($>10^8 M_{\Sun}$), and that the orbit of the progenitor does not follow the tidal debris, particularly in distance.  However, small tweaks to the orbit that make the simulations more consistent with the data change the orbit parameters by 10\% or less.  With the data presented here, we are unable to constrain the Galactic gravitational potential; further data on the CPS (or combining the data with other tidal streams) will be necessary to provide insight into the shape and strength of the halo potential probed by the stream.

In Section~\ref{sec:position} we noted that we were unable to separate the CPS from Sagittarius using SDSS photometry of F turnoff stars. In RGB stars, which are common to both
populations, we can only see the CPS if we have spectra, and can
separate the CPS stars by velocity and metallicity.  To study halo
substructure in more detail, we require spectroscopic surveys that
reach F turnoff stars at $g$ magnitudes approaching 22.
Alternatively, accurate proper motions could be used to separate the
CPS from the Sgr tidal stream, because the two streams are orbiting in opposite
directions, as was pointed out by
\citet{kbe+12}.  Our orbit fit predicts mean proper motions in the longitude bins with $\langle b \rangle \approx -66^\circ, -59^\circ, -53^\circ, -46^\circ,$ and $-36^\circ$ (see Tables~\ref{cpsk} and \ref{tab:cpsnbody}) of $(\mu_\alpha \cos{\delta}, \mu_\delta) \approx (1.2, -0.2), (1.3, -0.2), (1.3, -0.1), (1.4, -0.1), (1.5, 0.1)$~mas~yr$^{-1}$, respectively.
We were unable to distinguish the F turnoff stars in the two streams using currently available proper motions, though it might
be possible to separate larger populations of faint F turnoff stars
using future proper motion surveys. It is unlikely that the proper motion
measurement of individual stars will accurate enough, but it has been shown \citep{cmc+12,kbe13} that the ensemble proper motion of turnoff stars in the Sgr stream at distances of $\sim30-40$~kpc can be statistically determined from currently available data.

%.r orbitplots_cetus
%orbitplots,'orb.cps_bestfit.dat',tlim=1.5
%;-70<b<-62: <mu_alpha>=       1.2389063, <mu_delta>=     -0.22454619
%;-62<b<-56: <mu_alpha>=       1.2792962, <mu_delta>=     -0.17912219
%;-56<b<-50: <mu_alpha>=       1.3290676, <mu_delta>=     -0.12548441
%;-50<b<-42: <mu_alpha>=       1.3936136, <mu_delta>=    -0.056586964
%;-42<b<-30: <mu_alpha>=       1.5072909, <mu_delta>=     0.063549570
%
%

We simulate the evolution of satellites on the CPS orbit in a Milky Way-like potential via $N$-body modeling. From these mass-follows-light models, we show that with a mass-follows-light assumption, a $10^8 M_{\Sun}$ satellite can produce a stream width, line-of-sight velocity, and velocity dispersion similar to our observations. However, the distribution of stars along the stream and the apparent number of stars observed favor a lower mass satellite. It is possible to reconcile these observations if the dwarf galaxy was deflected into its current orbit on the order of one Gyr ago (less than one and a half orbits).  However, a less contrived solution is that mass does not (or did not) follow light in the CPS progenitor, which may have been a dark matter-dominated satellite similar to the recently-discovered class of satellites known as ``ultra-faint dwarf spheroidals" (e.g., Simon \& Geha 2007, McConnachie 2012). These are low luminosity, highly dark matter-dominated satellites. The mean metallicity, $\langle{\rm [Fe/H]} \rangle \approx -2.2$, that we find for CPS debris is consistent with the typical metallicities of ultra-faint dwarfs. More detailed modeling should be carried out in the future to confirm that dual-component (dark matter + stars) satellites reproduce the observed properties of the Cetus Polar Stream, and that the properties of the progenitor can be determined uniquely.  Our results suggest that the structure of tidal streams could be used to constrain the properties of the stream progenitors, including their dark matter components.

\acknowledgements

We thank the anonymous referee for thoughtful and detailed comments that greatly improved this work. 
This work was supported by the National Science Foundation grant AST
10-09670.  JLC was supported by NSF grant AST 09-37523. WY was
supported by NSF grant AST 11-15146. EO was supported by NSF grant DMR
08-50934.  JD was supported by the NASA/NY Space Grant fellowship.

Funding for SDSS-III has been provided by the Alfred P. Sloan
Foundation, the Participating Institutions, the National Science
Foundation, and the U.S.  Department of Energy Office of Science.  The
SDSS-III web site is http://www.sdss3.org/.

SDSS-III is managed by the Astrophysical Research Consortium for the
Participating Institutions of the SDSS-III Collaboration including the
University of Arizona, the Brazilian Participation Group, Brookhaven
National Laboratory, University of Cambridge, Carnegie Mellon
University, University of Florida, the French Participation Group, the
German Participation Group, Harvard University, the Instituto de
Astrofisica de Canarias, the Michigan State/Notre Dame/JINA
Participation Group, Johns Hopkins University, Lawrence Berkeley
National Laboratory, Max Planck Institute for Astrophysics, Max Planck
Institute for Extraterrestrial Physics, New Mexico State University,
New York University, Ohio State University, Pennsylvania State
University, University of Portsmouth, Princeton University, the
Spanish Participation Group, University of Tokyo, University of Utah,
Vanderbilt University, University of Virginia, University of
Washington, and Yale University.

\bibliographystyle{apj}

\end{document}